\documentclass[showpacs,preprintnumbers,amsmath,amssymb,eqsecnum]{revtex4}
\usepackage{graphicx}
\usepackage{subfigure}
\usepackage[dvips]{epsfig}
\usepackage{bm}
\usepackage{hyperref}
\def\cc{{\cal C}}

\def\II{{\cal I}}

\begin{document}


\title{Universality and stationarity of the I-Love relation for self-bound stars}

\author{T.~K. Chan\footnote{Present address: Department of Physics,
University of California at San Diego, 9500 Gilman Drive, La
Jolla, CA 92093, USA. Email address: chantsangkeung@gmail.com},
AtMa P.~O. Chan\footnote{Present address: Department of Physics,
University of Illinois at Urbana-Champaign,
 Urbana, IL 61801-3080, USA. Email address: atma.pochan@gmail.com}}
\author{P.~T. Leung\footnote{Email:
ptleung@phy.cuhk.edu.hk}} \affiliation{Physics Department and
Institute of Theoretical Physics, The Chinese University of Hong
Kong, Shatin, Hong Kong SAR, China.}

\date{\today}
\begin{abstract}
The emergence of  the  I-Love-Q  relations,
revealing that the moment of inertia, the tidal Love number
(deformability) and the spin-induced quadrupole moment of compact
stars are, to high accuracy,
 interconnected in a universal way disregarding the wide variety of
equations of state (EOSs) of dense matter, has attracted much
interest recently. However, the physical origin of these relations
is still a debatable issue. In the present paper, we focus on the
I-Love relation for self-bound stars (SBSs) such as incompressible
stars and  quark stars. We formulate perturbative expansions for
the moment of inertia, the tidal Love number (deformability) and
the I-Love relation of SBSs. By comparing the respective I-Love
relations of incompressible stars and a specific kind of SBSs, we
show analytically that the I-Love relation is, to relevant leading
orders in stellar compactness, stationary with respect to changes
in the EOS about the incompressible limit. Hence, the universality
of the I-Love relation is indeed attributable to the proximity of
compact stars to incompressible stars,
 and the stationarity of the relation as unveiled here.
We also discover that the moment of inertia and the tidal
deformability of a SBS with finite compressibility are, to leading
order in compactness, equal to their counterparts of an
incompressible star with an adjusted compactness, thus leading to
a novel explanation for the I-Love universal relation.
\end{abstract}
\maketitle

\section{Introduction}
The study of the static and dynamical properties of compact stars,
including both neutron stars (NSs) and quark stars (QSs), is often
plagued by the uncertainties  in the equation of state (EOS) of
nuclear (or quark) matter (see, e.g.,
\citep{Lattimer:2001,Lattimer:2005p7082,lattimer2007nso}), whose
stable state is currently unachievable in terrestrial
laboratories. On the other hand, such uncertainties also cast
doubt on the feasibility of testing the validity of Einstein's
theory of general relativity  with the structural characteristics
of compact stars because there is no a priori way to distinguish
between the effects due to the modifications in EOS and the theory
of gravity. Yet, the recently discovered ``I-Love-Q universal
relations" \citep{Yagi:2013long,Yagi:2013} reveal that the moment
of inertia $I$, the quadrupole tidal Love number $k_2$ (or tidal
deformability $\lambda_{ }$ \citep{Damour:09p084035,Yagi_14_PRD})
and the spin-induced quadrupole moment  $Q$ of compact stars
(including both NSs and QSs), after suitably scaled by the stellar
mass $M$, are inter-related in an almost EOS-independent way. The
discovery of the I-Love-Q relations is surprising and encouraging
as well. In the light of these universal relations, even still in
the absence of the exact  knowledge of nuclear (or quark) matter,
the possibility of identifying alternate theories of gravity
(e.g., the dynamical Chern-Simons gravity
\citep{Yagi:2013long,Yagi:2013}, the Eddington-inspired
Born-Infeld  gravity \citep{Sham_14_ApJ},  the scalar-tensor
theories
\citep{Pani_Berti2014,Doneva_14_apj,Kleihaus_PRD_14,Kokk_scalar_IQ}
and the $f(R)$ gravity theories \cite{Doneva:2015hsa})
 has been examined. In addition,
these relations were shown to be robust and prevailing in  other
stellar systems, such as close binary compact stars
\citep{Maselli:2013} and rapidly rotating compact stars (including
both unmagnetized and magnetized NSs)
\citep{Pappas_14_prl,Chak_14_prl,Haskell_14_mnras}, and their
generalizations to  higher-order multipole moments induced by
either tidal forces or rotation have been proposed
\citep{Yagi_14_PRD,Yagi_hair_GR,Stein_hair_apj}. Through  such
relations, the measurement of  either member of the  I-Love-Q trio
will suffice to determine the other two  once the mass of a
compact star is known \citep{Yagi:2013long,Yagi:2013}. As a
result, more accurate astronomical data can be extracted  in the
analysis of gravitational wave signals emitted at the late stage
of NS-NS binary mergers
\citep{Flanagan:08p021502,Hinderer:08p1216,Yagi:2013long,Yagi:2013}.

To certain extent, the I-Love-Q  relations have emerged as a
surprise to researchers working on compact stars, who are
accustomed to the fact that the physical characteristics of NSs
(or QSs), including all members of the I-Love-Q trio, usually bear
obvious dependence on  the EOS of nuclear (or quark) matter (see,
e.g.,
\citep{Lattimer:2001,Lattimer:2005p7082,lattimer2007nso,Damour:09p084035,Lattimer_Love}).
Yet the I-Love-Q trio are found to be related by universal
formulas which hold for both NSs and QSs, and are accurate up to a
few percent \citep{Yagi:2013long,Yagi:2013}. How can they conspire
to follow such EOS-insensitive formulas? In the seminal papers
discovering the  I-Love-Q relations,
\citet{Yagi:2013long,Yagi:2013} have put forward two suggestions
for the solution of this puzzle: (i) the low-density region of a
NS, which lies in a layer between $70 \%$ and $90 \%$ of  the
stellar radius $R$, contributes the most to the quantities $I$,
$\lambda$ and $Q$ and the EOS there
 is quite unified; and (ii) the I-Love-Q relations are indeed the remnants of the no-hair theorem
 of black-holes. However, the subsequent finding that QSs also follow the I-Love-Q relations \citep{Yagi_14_PRD,ILoveQ_1,ILoveQ_2}
 seems to invalidate suggestion (i) because the behavior of quark matter is completely different from that of nuclear matter
 in the low density regime. Furthermore, each of the I-Love-Q trio is
 in fact
dominated by a thicker layer bounded between $0.5R$ and $0.95R$
 \citep{whyI}, which comprises both high and
low-density matter. Hence, the universality demonstrated in the
I-Love-Q relations is unlikely to be rooted in the low-density
region of compact stars. On the other hand, the I-Love-Q relations
also hold nicely in the Newtonian limit \citep{Yagi:2013long}.
Thus, the no hair theorem for black-holes seems to be irrelevant
to these relations.

More recently, \citet{ILoveQ_1} revealed that the I-Love-Q
relations of realistic compact stars, whose effective polytropic
index is less than unity in the high density regime, indeed follow
closely those of incompressible stars (ISs). In particular, the
accuracies of these relations are found to deteriorate
significantly for polytropic stars with  polytropic indices
greater than 1. Moreover, \citet{ILoveQ_3} analytically derived
the I-Love relation for relativistic ISs and showed that the
formula also applies to realistic NSs and QSs with high
accuracies. Therefore, they proposed that  the I-Love-Q
universality is the consequence of (i) the high stiffness of dense
nuclear matter, and (ii) the stationarity of the I-Love-Q
relations about the incompressible limit
\citep{ILoveQ_1,ILoveQ_2,ILoveQ_3}.

The main objective of the present paper is to provide a firm
theoretical support to the above-mentioned observation that the
I-Love-Q relations is, to a specific order of stellar compactness
$\cc$, stationary with respect to variations in the
compressibility of nuclear (or quark) matter forming a compact
star about the incompressible limit (i.e., vanishing
compressibility). While the said issue has been verified
qualitatively by numerical data shown in
\citep{ILoveQ_1,ILoveQ_3}, its theoretical justification is not
yet fully  known. \citet{ILoveQ_1} has proposed a generalized
Tolman model (GTM), whose density profile $\rho(r)$ is given by
\begin{equation}\label{GTM}
\rho(r)=\rho_c (1- \delta x^2),
\end{equation}
with $\rho_c$ being the central density, $x \equiv r/R$, $0 \le
\delta \le 1$. The parameter $\delta$  vanishes for ISs and, in
general, measures the compressibility of a GTM star
\citep{ILoveQ_2}. For example, near the stellar center, GTM can
mimic  the density distribution of a polytropic star with
polytropic index $N \simeq \delta$. For such GTM stars,
\citet{ILoveQ_1} show that the I-Love-Q relations are stationary
with respect to changes in $\delta$ about the point $\delta=0$
(i.e., ISs) in the Newtonian limit. However, whether similar
behaviors still prevail in relativistic stars constructed with
realistic EOSs is not yet clearly examined theoretically.

In the present paper, we intend to develop a fully relativistic
approach to the study of such universality. Our focus is the
stationarity of the I-Love relation for self-bound stars (SBSs),
including both ISs and QSs \citep{Johnson,SS,Witten} (see
\eqref{QS_AS1} for the EOS of SBSs), for their proximity to ISs.
As demonstrated numerically by \citet{ILoveQ_3}, the I-Love
formulas for ISs and QSs are almost the same. Here we will carry
out an in-depth analytic study on such similarity between these
two classes of SBSs. Our target is to pinpoint the physical
mechanism underlying the universality of the I-Love relation.

First of all, in order to evaluate the moment of inertia and the
tidal Love number (or deformability), we need to find the
hydrostatic equilibrium configuration   of relativistic stars,
which is governed by the Tolman-Oppenheimer-Volkov (TOV) equations
\cite{Oppenheimer, Tolman}. To this end, we develop a recursive
post-Minkowsian scheme to solve for the density, the pressure
 and the metric coefficients of SBSs. Each of these
physical quantities is shown to be expressible in terms of a power
series in stellar compactness, which is then used as the input to
evaluate the moment of inertia and the tidal Love number (or
deformability) of SBSs.

With the abovementioned expansion scheme,  we succeed in finding
the post-Minkowsian expansion for the moment of inertia and the
Love number of QSs obeying the simple MIT bag model  (see, e.g.,
\citep{Johnson,SS,Witten} and \eqref{QMEOS}). Both of these two
quantities  are expressed in terms of power series in compactness.
We combine these two series to express the scaled moment of
inertia as a power series in $\bar{\lambda}^{-1/5}$, where
$\bar{\lambda}$ is the dimensionless tidal deformability
\citep{Damour:09p084035,Lattimer_Love,Yagi:2013long,Yagi:2013,Yagi_14_PRD}
and $\bar{\lambda}^{-1/5}$ is, to leading order, proportional to
the stellar compactness $\cc$. Thus, the I-Love relation of  QSs
is found. Comparing the I-Love relation for  QSs with that of ISs,
which has been obtained recently by \citet{ILoveQ_3}, we find that
they are identical up to first-order in $\bar{\lambda}^{-1/5}$.
Therefore, in spite of the difference in their  EOSs, QSs and ISs
actually obey the same I-Love relation up to first-order in
compactness.
 As a result, the
I-Love relations for QSs and ISs exactly coincide in the Newtonian
limit.

Moreover, we consider in general a class of SBSs characterized by
 EOS  $\rho =c_0+c_1 p$, which is a linear function of pressure $p$ and reduces to the MIT bag model
if the two positive parameters $c_0 $ and $c_1 $ are
suitably chosen. 
Again we express the I-Love relation for such (linear) SBSs as
Taylor series in  $\bar{\lambda}^{-1/5}$.  We show that the I-Love
relation is completely unaffected by the value of $c_0$. More
importantly, to first-order in compactness such relation is also
independent of $c_1$, which is indeed the inverse of the square of
the sound speed and a measure of the compressibility. As the
 compactness of typical compact stars is usually less than 0.3,
 the influence of $c_1$ on the I-Love relation is expected to be
 small. This point is indeed verified numerically in the QS case.
 The analysis developed here consequently provides a strong
theoretical justification to the stationarity of the I-Love
relation about the incompressible limit as discovered in
\citep{ILoveQ_1,ILoveQ_3}.

Finally, we generalize our study to SBSs with EOSs given by
regular power series in pressure $p$ (see \eqref{QS_AS1}) and show
that, to respective leading orders in stellar compactness, the
I-Love relation is still stationary with respect to variations in
all the expansion coefficients of the EOS about the incompressible
limit. Thus, the cause for the universality of the I-Love relation
of realistic stars, which are characterized by sufficiently stiff
EOS, is is fully exposed. In addition, we also provide a
physically transparent explanation for such stationarity, which is
attributable to (i) the similarity  between the responses of the
moment of inertia and the tidal Love number to changes in EOS, and
(ii) the proper scaling in the definition of the two variables
(i.e., the scaled moment of inertia $\bar{I} \equiv I/M^3$ and the
dimensionless tidal deformability $\bar{\lambda}$) considered in
the I-Love relation. We show that to leading order of compactness
$\cc$, the changes in $\bar{I}$ and $\bar{\lambda}$ induced by a
non-zero compressibility of stellar EOS are reproducible by a
corresponding shift in the compactness of ISs. As a result, in
spite of the fact that both $\bar{I} - \cc $ and $\bar{\lambda}
-\cc$ relations display obvious EOS-dependency, the $\bar{I}-
\bar{\lambda} $ relation obtained from eliminating $\cc$ in these
two relations is approximately independent of EOSs.

The paper is organized as follows. In Section \ref{SBS} a
recursive perturbation scheme is established to solve the TOV
equations \cite{Oppenheimer, Tolman} for SBSs. Various physical
quantities describing the equilibrium configuration  of SBSs (such
as density, pressure and metric coefficients) are expanded in
terms of power series in compactness. With these expansions, in
Sections \ref{moment} and \ref{Love}  we formulate the
perturbative expansions for the moment of inertia and the Love
number for SBSs, respectively. In Section \ref{I_Love} we find an
analytic formula relating the moment of inertia and the Love
number for SBSs described by EOS linear in pressure. In
particular, we compare the I-Love relations of ISs and QSs (or
other SBSs characterized by linear EOS) and show that the
relations in these two cases are identical to first-order in
stellar compactness. In Section~\ref{general} we consider the
I-Love relation of SBSs with EOSs depending on the pressure
nonlinearly and show that the relation is, to relevant orders of
stellar compactness, still stationary to changes in all the
parameters specifying such nonlinear EOSs about the incompressible
limit. As a result, the robustness of the stationarity property of
the I-Love relation is established in general. In
Section~\ref{physics} we further discuss and elaborate the
physical origin of such stationarity by comparing the respective
changes in the scaled moment of inertia $\bar{I}$ and the
dimensionless tidal deformability  $\bar{\lambda}$ due to a
specific change in EOS.
 We conclude our paper in Section \ref{Conclusion} with
some discussions. Besides, for readers who are interested in the
accuracies of the post-Minkowsian expansions developed here, we
summarize the relevant numerical results for the QS case in
Appendix~A of our paper.
 Unless otherwise
stated explicitly, geometric units in which $G=c=1$ are adopted.

\section{Hydrostatic equilibrium  of SBS}\label{SBS}
\subsection{TOV equations}\label{TOV}
The hydrostatic equilibrium  of
 a relativistic, non-rotating compact star made of  a perfect fluid
 is governed by
the   TOV equations \cite{Oppenheimer, Tolman}:
\begin{eqnarray}
\frac{dp}{dr}&=&-\frac{(m+4\pi r^{3}p)(\rho+p)}{r^{2}(1-2m/r)}~,\label{TOV1}\\
\frac{d\nu}{dr}&=&-\frac{2}{\rho+p}\frac{dp}{dr}~,\label{TOV2}
\end{eqnarray}
where  $\rho(r)$, $p(r)$  and  $e^{\nu(r)}$ are the energy
density, pressure and the metric coefficient at a circumferential
radius $r$, respectively, and $m(r)= 4\pi\int_0^r \rho (r') r'^2
dr' $ is the gravitational mass enclosed within radius $r$
\cite{Oppenheimer, Tolman}. The pressure vanishes at the stellar
surface where $r=R$. Outside the star, the spacetime metric is
given by the Schwarzschild metric
\begin{equation}\label{StaticMetric}
ds^{2}=-e^{\nu (r)}dt^{2}+e^{\lambda
(r)}dr^{2}+r^{2}\left(d\theta^{2}+\sin^{2}\theta d{
\phi}^{2}\right)~,
\end{equation}
where $t$, $r$, $\theta$, ${ \phi}$ are the  standard
Schwarzschild coordinates, and $e^{\nu}=e^{-\lambda}=1-2M/r$ (see,
e.g., \cite{Hartle, ShapiroComstar}). Inside the star, while
$e^{\lambda (r)}$ is directly given by
$e^{\lambda}=1/[1-2m(r)/r]$, $e^{\nu (r)}$ has to be obtained by
solving the TOV equations listed above subject to  the boundary
condition $e^{\nu(R)}=1-2M/R$.

\subsection{Series solution of TOV equations for SBSs}\label{SBSP}
A SBS is composed of self-bound material whose energy density is
greater zero even at zero pressure (see, e.g.,
Ref.~\citep{Lattimer_Love}). Throughout the present paper we
assume that its energy density $\rho$ is expressible in terms of a
regular Taylor series in pressure $p$:
\begin{equation}\label{QS_AS1}
\rho=c_{0}+c_{1}p+c_{2}p^{2}+\cdots~,
\end{equation}
where $c_n$ ($n=0,1,2,\cdots$) are real parameters characterizing
the EOS. We also assume that the above EOS can be truncated at a
certain order with good accuracy and
 $\rho\geq0$ for
any given $p \ge 0$. Physically speaking, $c_{0}>0$  and $c_{1}=(d
\rho /d p)_{p=0} \ge 0$ are, respectively, the energy density and
a measure of the compressibility  of the stellar matter at zero
pressure. In particular, for ISs $c_n=0$, where $n=1,2,3,\cdots$.
On the other hand, for QSs obeying the simple linear MIT bag model
(see, e.g., \citep{Johnson,SS,Witten}), $c_n=0$ for $n>1$.
However, if other factors (such as the mass of strange quarks and
finite temperature effect) are also considered, the EOS of QSs can
still be approximated by  a finite polynomial as given by
\eqref{QS_AS1} (see, e.g., \citep{SMatter, Q_star_T,ComStar,
haensel2007nse,Weber-pulsar}).

In order  to work out the I-Love relation for a SBS whose EOS is
given by \eqref{QS_AS1}, in the following discussion we formulate
a perturbative solution for the TOV equations. First of all, we
expand $p$, $\rho$, $\II \equiv m(r)/R^3$ and $\hat{\II} \equiv
\II(r=R)=M/R^3$ in terms of Taylor series in stellar compactness
$\cc \equiv M/R$ as follows:
\begin{eqnarray}
p(x;\cc)&=& p_{0}(x)+p_{1}(x)\cc+p_{2}(x)\cc^{2}+\cdots~,\label{p-exp}\\
\rho(x;\cc)&=&\rho_{0}(x)+\rho_{1}(x)\cc+\rho_{2}(x)\cc^{2}+\cdots~,\label{rho-exp}\\
\II(x;\cc)&=&\II_{0}(x)+\II_{1}(x)\cc+\II_{2}(x)\cc^{2}+\cdots~,\label{I-exp}\\
\hat{\II}(\cc)&=&\hat{\II}_{0}+\hat{\II}_{1}\cc+\hat{\II}_{2}\cc^{2}+\cdots~,\label{Ihat-exp}
\end{eqnarray}
where $x=r/R$ is the normalized radial coordinate,  $p_{n}(x)$,
$\rho_{n}(x)$ and  $\II_{n}(x)$ are functions of $x$, and
$\hat{\II}_{n} = \II_{n}(x=1)$ are constant coefficients for
$n=0,1,2,\cdots$.

  In terms of $\II$ and
$\hat{\II}$,  the TOV equation (\ref{TOV1}) can be rewritten as:
\begin{equation}\label{QS_AS2}
\frac{dp}{dx}=-\cc \frac{\left(\II+4\pi x^3
p\right)\left(\rho+p\right)}{x\left(x\hat{\II}-2\cc\II\right)}~.
\end{equation}
By expanding both sides of \eqref{QS_AS2} into power series of
$\cc$ and taking EOS \eqref{QS_AS1} into account, the functions
$p_{n}(x)$ and $\rho_{n}(x)$ can be solved recursively. The
explicit results for $p(x;\cc)/c_0$ and $\rho(x;\cc)/c_0$ up to
$\cc^2$-term are listed below for illustration:
\begin{eqnarray}
\frac{p(x;\cc)}{c_{0}}&=&\frac{\cc}{2}\left(1-x^{2}\right)
+\frac{\cc^{2}}{5}\left(1-x^{2}\right)\left[\left(5+c_{1}\right)-c_{1}x^{2}\right]
+\cdots~,\label{QS_AS8} \\
\frac{\rho(x;\cc)}{c_{0}}&=&1+\frac{\cc
}{2}\left(1-x^{2}\right)c_{1}
+\frac{\cc^{2}}{20}\left(1-x^{2}\right)\left[\left(20c_{1}+4c_{1}^{2}+5c_{0}c_{2}\right)-\left(4c_{1}^{2}+5c_{0}c_{2}\right)
x^{2}\right] +\cdots~,\label{QS_AS9}
\end{eqnarray}
and the expansions for $\II(x;\cc)$ and $\hat{\II}(\cc)$ follow
directly from their definitions. We note here that (i) in general
both $p_{n}(x)$ and $\rho_{n}(x)$ are $n$-th degree polynomials in
$x^2$; and (ii) $\rho(x;\cc)$ tends to $c_0$ in the
zero-compactness limit; and (iii) to first-order in $\cc$,
$\rho(x;\cc)$ is in fact given by the GTM proposed in
Refs.~\citep{ILoveQ_1,ILoveQ_2}.

Likewise, the expansion for the metric coefficient $e^{\nu}\equiv
\sum^\infty_{n=0}(e^{\nu})_{n}(x)\cc^n$ can also be found using
the TOV equation~(\ref{TOV2}):
\begin{eqnarray}
e^{\nu}(x;\cc) =1-\left(3-x^{2}\right)\cc
+\frac{3}{20}(5-c_{1})\left(1-x^{2}\right)^{2}\cc^{2}+\cdots~.
\label{QS_AS10}
\end{eqnarray}
It agrees with the Minkowskian metric in the zero-$\cc$ limit and,
remarkably, is independent of $c_0$. Besides, to first-order in
$\cc$, $e^{\nu}$ also does not depend on the EOS of the self-bound
matter. Hence, the metric profile $e^{\nu}$ for SBSs is universal
in the low-compactness regime.

Lastly,  the mass-radius ($M$-$R$) relation of SBSs can also found
by eliminating $\cc$ from the following pair of relations
\begin{eqnarray}
 M&=&\cc R ~, \label{mass} \\ R&=&\left[\frac{\cc}{\hat{\II}(\cc)}\right]^{1/2}~, \label{radius}
\end{eqnarray}
 resulting in
\begin{equation}\label{MRR1}
M=\frac{4}{3}\pi c_{0}R^{3}\Big[1+\frac{1}{5}c_{1}\hat{\cal
I}_{0}R^{2}+\frac{1}{35}\left(14c_{1}+3c_{1}^{2}+2c_{0}c_{2}\right)\hat{\cal
I}_{0}^{2}R^{4}+\cdots\Big]~,
\end{equation}
with $\hat{\cal I}_{0} = 4\pi c_0/3$. Note that the first term in
the above equation  corresponds to the Newtonian result for ISs,
while other higher order terms account for the general
relativistic corrections
 due to the compressibility of the
self-bound matter. We have verified the validity and gauged the
accuracies of the perturbative expansions obtained in this section
by applying them to QSs described by the simple MIT bag model.
While the detailed result of such  investigation can be found in
Ref.~\citep{atma}, we also include a short summary of the relevant
information in Appendix A.



\section{Moment of Inertia}\label{moment}
The moment of inertia $I$ of SBSs in the slowly rotating limit
(see, e.g., \cite{Hartle_1967,Hartle_1968,ComStar}) can be
obtained from the perturbative solution for the stellar
equilibrium configuration derived in Section~\ref{SBS} as follows.
First of all, the angular velocity of the local inertial frame,
$\Lambda(x)$, due to the frame-dragging effect of a compact star
(e.g., a SBS) rotating uniformly at a unit angular velocity
 is governed by the differential equation \cite{Hartle_1967, Hartle_1968}:
\begin{equation}\label{moi1}
\frac{d}{dx}\Big(x^{4}j\frac{d\Lambda}{dx}\Big)+4x^{3}\frac{dj}{dx}(\Lambda-1)=0~,
\end{equation}
where $0 \le x \le 1$ and $j(x)=e^{-(\lambda+\nu)/2}$.  Outside
the rotating star, where  $x>1$, $\Lambda(x)$ satisfies another
differential equation
\begin{equation}\label{outside}
\frac{d}{dx}\Big(x^{4}\frac{d\Lambda}{dx}\Big)=0~,
\end{equation}
which is much simpler and
 can be readily integrated to yield the result
$\Lambda=2I/r^3$ (see \cite{Hartle_1967,Hartle_1968} for the
details). Hence, the normalized moment of inertia $a \equiv
I/MR^{2}$ is given by $a=\hat{\Lambda}/(2\cc)$ with $\hat{\Lambda}
\equiv \Lambda(x=1)$ being the surface value of $\Lambda$.

In the following we generalize  the perturbative scheme proposed
recently by \citet{ILoveQ_3}, which was originally targeted at the
 moment of inertia of ISs, to find the moment of
inertia of SBSs. Firstly, it follows directly from (\ref{moi1})
and the regularity boundary condition of $\Lambda$ at $x=0$ that
\begin{equation}\label{moi3}
\frac{d\Lambda}{dx}=-\frac{1}{x^{4}j}\int_{0}^{x}4x'^{3}\frac{dj}{dx'}(\Lambda-1)
dx'~.
\end{equation}
In order to solve (\ref{moi3}) for $0 \le x \le 1$,  we expand
$\Lambda(x)$ and $j(x)$ in power series of $\cc$,
\begin{eqnarray}
\Lambda(x;\cc)&=&\Lambda_{0}(x)+\Lambda_{1}(x)\cc+\Lambda_{2}(x)\cc^{2}+\cdots~,\label{moi2}\\
j(x;\cc)&=&j_0(x)+j_1(x)\cc+j_2(x)\cc^{2}+\cdots~,
\end{eqnarray}
where $j_n(x)$, $n=0,1,2,\cdots$, can be found directly from the
expansions of $e^\lambda$ and $e^\nu$ obtained in
Section~\ref{SBS}. Substituting these expansions into \eqref{moi3}
and noting that (i) in the Newtonian limit
\begin{eqnarray}
\lim_{\cc \rightarrow 0} j(x ) &=& j_0(x)=1~;\\
\lim_{\cc \rightarrow 0} \Lambda(x ) &=& \Lambda_0(x)=0~;
\end{eqnarray}
and (ii)  $\Lambda_{n+1}=-(d\Lambda_{n+1}/dx)/3$ holds at $x=1$,
which is the consequence of the  continuity $\Lambda$ and
$d\Lambda/dx$ across the stellar surface,
 we can recursively find $\Lambda_{n+1}$ from
$\Lambda_{0},\Lambda_{1},\ldots,\Lambda_{n}$, and in turn arrive
at the expansion of
 $\Lambda$ (see \citep{ILoveQ_3} for the details):
\begin{equation}\label{moi5-1}
\Lambda(x;\cc)=\frac{2\cc}{5}\left(5-3x^{2}\right)
+\frac{\cc^{2}}{350}\left[35(3+c_{1})+126(5-c_{1})x^{2}-15(33-5c_{1})x^{4}\right]+\cdots~,
\end{equation}
which  holds generally for a uniformly rotating SBS in the slow
rotation limit. It can be noted from (\ref{moi5-1}) that
$\Lambda_{n}$ is in general an $n$-th degree polynomial in $x^2$.

With this analytic expression for $\Lambda(x;\cc)$, we can find
the normalized moment of inertia $a$ for an arbitrary SBS:
\begin{equation}
a(\cc)=\frac{2}{5}\Big[
1+\frac{2\cc}{35}(15-c_{1})+\frac{2\cc^{2}}{1575}\left(795-80c_{1}-7c_{1}^{2}-20c_{0}c_{2}\right)+\cdots\Big]~.\label{a:exp}
\end{equation}
In the zero-$\cc$ limit, $a=2/5$, which is just the Newtonian
result of a  uniform sphere.  With the expressions for $R$, $M$
and $a$ given respectively in (\ref{mass}), (\ref{radius})  and
\eqref{a:exp}, the moment of inertia $I=aMR^{2}$ of a SBS can be
straightforwardly found. On the other hand, in the I-Love-Q
relations, the scaled moment of inertia $\bar{I} \equiv
I/M^3=a/\cc^{2}$ is considered \citep{Yagi:2013long,Yagi:2013}.
Hence, Eq.~\eqref{a:exp} actually provides a simple way to find
$\bar{I}$ perturbatively for SBSs.

\section{Tidal deformation}\label{Love}
In this section we outline the method of evaluating the tidal Love
number (or deformability) of SBSs from the stellar equilibrium
configuration obtained Section~\ref{SBS}. Consider a non-rotating
compact star acted on by an external tidal field $E_{ij}$
\cite{Damour:09p084035,Hinderer:08p1216}. As a result, the star
acquires a quadrupole moment $Q_{ij}$ given by
\begin{align}
Q_{ij}=-k_2\left(\frac{2R^5}{3}\right)E_{ij}\equiv -\lambda
E_{ij}~,
\end{align}
where $k_2$ (dimensionless) and $\lambda$ are called the tidal
Love number and deformability, respectively. In the I-Love-Q
relations, the dimensionless  tidal deformability
$\bar{\lambda}\equiv\lambda/M^5=2k_2/(3\cc^5)$ is considered
\citep{Yagi:2013long,Yagi:2013}.

We follow the formulation developed  in
\cite{Damour:09p084035,Hinderer:08p1216,Lattimer_Love} to evaluate
the tidal Love number $k_2$. First of all, the logarithmic
derivative of the metric perturbation
 $H=H_0=H_2$
 (see \citep{Price,Lindblom-1997} for the conventions of the metric
 functions), $y \equiv rH'(r)/H(r)$, is considered. The
following nonlinear ordinary differential equation governs the
variation of  $y(r)$ inside the star
\cite{Damour:09p084035,Hinderer:08p1216,Lattimer_Love}:
\begin{align}
\label{ylove-1} r
\frac{dy(r)}{dr}+y(r)^2&+y(r)e^{\lambda(r)}\left\{1+4\pi
r^2[p(r)-\rho(r)]\right\}+r^2Q(r)=0,
\end{align}
where
\begin{align}
Q(r)=&4\pi
e^{\lambda(r)}\left[5\rho(r)+9p(r)+\frac{\rho(r)+p(r)}{c_s^2(r)}\right]
-6\frac{e^{\lambda(r)}}{r^2}-\left[\frac{d\nu(r)}{dr}\right]^2~,
\end{align}
$c_s \equiv \sqrt{d p/d \rho}$ is the speed of sound, and
$y(r=0)=2$. After solving \eqref{ylove-1} inside the star to
evaluate the surface value of the logarithmic derivative of the
metric perturbation, $y_R \equiv y(R^+) = y(R^-)-4 \pi R^3 \rho
(R^-)/M$, the tidal Love number $k_2$ can then be found from the
following formula
\cite{Damour:09p084035,Hinderer:08p1216,Lattimer_Love}:
\begin{align}
\label{k2eq}
k_2({\cal C},y_R)=&\frac{8}{5}{\cal C}^5 (1-2{\cal C})^2[2{\cal C}(y_R-1)+2]\left\{2{\cal C}[4(y_R+1){\cal C}^4+(6y_R-4){\cal C}^3+(26-22y_R){\cal C}^2+3(5y_R-8)-3y_R+6]\right.\nonumber\\
&\left.+3(1-2{\cal C})^2[2{\cal C}(y_R-1)-y_R+2]\log(1-2{\cal C})
\right\}^{-1}~.
\end{align}

In order to solve the nonlinear differential equation
\eqref{ylove-1} for SBSs, as suggested recently by
\citet{ILoveQ_3}, we substitute the following power series
expansion of $y$,
\begin{align}
\label{yexp} y(x)=y_0(x)+y_1(x){\cal C}+y_2(x){\cal
C}^2+y_3(x){\cal C}^3+\cdots~,
\end{align}
and the series expansions of other relevant physical quantities
obtained previously into the equation. By matching the
coefficients of equal powers of ${\cal C}$  in the resultant
equation, we find a set of coupled linear first-order ordinary
differential equations \citep{ILoveQ_3}:
\begin{eqnarray}
\label{firsty} &&x y_0'(x)+y_0(x)^2+y_0(x)-6=0, \\
\label{2-y}&&x
y_1'(x)+[1+2y_0(x)]y_1(x)-x^2y_0(x)+3(c_1+1)x^2=0,~\cdots,
\end{eqnarray}
which can be solved recursively by imposing the boundary
conditions $y_0(0)=2$ and $y_n(0)=0$ for $n=1,2,\cdots$ as
follows:
\begin{eqnarray}
y_0(x)&=&2;\;\;\;\\
y_1(x)&=&-\frac{3c_1+1}{7}x^2;
\end{eqnarray}
etc. From these  solutions we can then find $y_R$, and hence $k_2$
from \eqref{k2eq}:
\begin{eqnarray}
\label{k2-general} k_2& =&(1-2\cc)^2 [\frac{3}{4} -
\frac{3\cc}{28} (12 + c_1) +
 \frac{\cc^2}{2940} (1210 - 230 c_1 - 37 c_1^2 - 140 c_0 c_2)
    +\cdots]~.
\end{eqnarray}

\section{I-Love relation for linear EOS}\label{I_Love}
In the present paper, we aim to understand why the I-Love relation
is so insensitive to variation in the EOS as long as the star in
consideration is sufficiently stiff. To this end, we compare two
representative members of SBSs with EOSs given by \eqref{QS_AS1},
namely, ISs and QSs. It is obvious that ISs are the simplest case
of SBSs with $c_{n}=0$ for $n\geq 1$ and $c_{0}$ being the
constant density of a star. Of course, ISs are also the stiffest
stars and, as shown in \citep{ILoveQ_3}, their I-Love relation can
accurately approximate those of realistic stars. On the other
hand, in the MIT bag model for the EOS of quark matter
\citep{Johnson,SS,Witten}, if the effects of non-zero quark masses
and temperature are omitted, the EOS then takes the linear form:
\begin{equation}\label{QMEOS}
\rho=4B+3p~,
\end{equation}
where $B>0$ is called the bag constant. It is a special case of
SBSs with $c_{0}=4B$, $c_{1}=3$ and $c_{n}=0$ for $n\geq2$. As
verified numerically in \citep{ILoveQ_1,ILoveQ_3}, the I-Love
relations of ISs and QSs are almost identical, especially in the
Newtonian limit. Here we look into the interrelationship between
the I-Love relations of these two kinds of SBSs with the
perturbative scheme developed in the previous sections.

Using the abovementioned  series expansion method for ISs and QSs,
we express the scaled moment of inertia $\bar{I}$ and the
dimensionless tidal deformability $\bar{\lambda}$ in terms of the
stellar compactness $\cc$. Moreover, we can also eliminate $\cc$
from the expressions for $\bar{I}$ and $\bar{\lambda}$ and relate
$\bar{I}$ to $\bar{\lambda}$ directly. The results for ISs are
summarized as follows:
\begin{eqnarray}
\bar{I}&=&\frac{2}{5\cc^2}+\frac{12}{35\cc}+\frac{212}{525}+
\frac{632}{1155}{\cal C}+\frac{703744}{875875}{\cal
C}^2+\frac{251264}{202125}{\cal C}^3+
\frac{121542272}{60913125}{\cal C}^4+\cdots~.\label{IS-I} \\
\bar{\lambda}_\cc &=&
\frac{3}{4}-\frac{9 {\cal C}}{7}+\frac{121 {\cal C}^2}{294}-\frac{479
   {\cal C}^3}{11319}-\frac{196375 {\cal C}^4}{1030029}-\frac{10670812
   {\cal C}^5}{21630609}-\frac{32621700682 {\cal
   C}^6}{28314467181}+\cdots
   ,\label{IS-k2}\\
 \bar{I}&=& \frac{2}{5 \zeta^2}+\frac{44}{35
\zeta}+\frac{17452}{11025}+\frac{31936
}{33957}\zeta+\frac{21242792 }{105343875}\zeta^2-\frac{990746384
}{24334435125}\zeta^3-\frac{59041871509888}{1433419901038125}\zeta^4
+\cdots~,\nonumber  \\
&=&
\bar{\lambda}^{2/5}\left(0.5278+\frac{1.444}{\bar{\lambda}^{1/5}}+\frac{1.583}{\bar{\lambda}^{2/5}}
+\frac{0.8187}{{\bar{\lambda}^{3/5}}}
+\frac{0.1528}{{\bar{\lambda}^{4/5}}}-\frac{0.02686}{{\bar{\lambda}}}-\frac{0.02366}{{\bar{\lambda}^{6/5}}}
+\cdots \right)~\label{IS-ILove}.
\end{eqnarray}
where $\zeta \equiv (2\bar{ \lambda})^{-1/5}$ and, for
convenience, the dimensionless compactness-scaled tidal
deformability
\begin{equation}\label{scaled_l}
 \bar{\lambda}_\cc \equiv \frac{3 \cc^5\bar{\lambda}}{2(1-2{\cal
 C})^2}=\frac{k_2}{(1-2{\cal
 C})^2}
\end{equation}
is introduced. We note that (i) both $\bar{I}$ and $\bar{\lambda}$
are independent of the density of the star in these equations, and
(ii) the above results are identical to what have been obtained in
\citep{ILoveQ_3} using the Schwarzschild constant-density solution
for the TOV equations \citep{Tolman_book} as the input to evaluate
the moment of inertia and tidal Love number.

On the other hand, we can similarly find the associated formulas
for QSs:
\begin{eqnarray}
\bar{I}&=&\frac{2}{5 {\cal C}^2}+\frac{48}{175{\cal C}
}+\frac{656}{2625}+\frac{40408}{202125}{\cal C}-\frac{883424
}{21896875}{\cal C}^2-\frac{24137984 }{25265625}{\cal
C}^3-\frac{339641200144 }{83755546875}{\cal C}^4+\cdots,\label{QS-I}\\
\bar{\lambda}_\cc&=&
\frac{3}{4}-\frac{45 {\cal C}}{28}+\frac{187
   {\cal C}^2}{2940}-\frac{1141589 {\cal C}^3}{1131900}-\frac{273911383
   {\cal C}^4}{103002900}-\frac{374294273707
   {\cal C}^5}{54076522500}
   -\frac{6469553810716163
   {\cal C}^6}{353930839762500}+\cdots,
   \label{QS-k2}\\
\bar{I}&=&+\frac{2}{5 \zeta^2}+\frac{44}{35
\zeta}+\frac{87134}{55125}+\frac{1783052
   \zeta}{1929375}+\frac{181653418 \zeta^2}{1158782625}-\frac{1797421853576 \zeta^3}{15209021953125}
   -\frac{27761427683880668 \zeta^4}{179177487629765625}+\cdots,\nonumber\\
   &=&\bar{\lambda}^{2/5}\left( 0.5278 +\frac{1.444}{\bar{\lambda}^{1/5}}+\frac{1.581}{\bar{\lambda}^{2/5}}
   +\frac{0.8045}{\bar{\lambda}^{3/5}}+\frac{0.1188}{\bar{\lambda} ^{4/5}}-\frac{0.07797}{\bar{\lambda}}
   -\frac{0.08899}{\bar{\lambda}^{6/5}}+\cdots \right).
   \label{QS-ILove}
\end{eqnarray}
It is remarkable that in all these expansions for ISs (or QSs) the
dependency on $c_0$ (or the bag constant $B$) disappears.

\begin{figure}
\centering \caption{(Upper panel) $\log_{10}\bar{I}$ is plotted
against $\log_{10}\bar{\lambda}$  for ISs and QSs. The numerical
data (circles for ISs and squares for QSs) are compared with the
analytical results obtained from  the 7-term post-Minkowskian
expansions \eqref{IS-ILove} (for ISs, the continuous line) and
\eqref{QS-ILove} (for QSs, the dashed line). (Lower panel) The
logarithm of the relative error between the numerical data and
series expansion results, $E \equiv |(\bar{I})_{\rm series}/(
\bar{I})_{\rm data}-1 |$, is shown against
$\log_{10}\bar{\lambda}$
 (with circles for ISs and squares for QSs).
}\label{fig-error}
\includegraphics[scale=0.8]{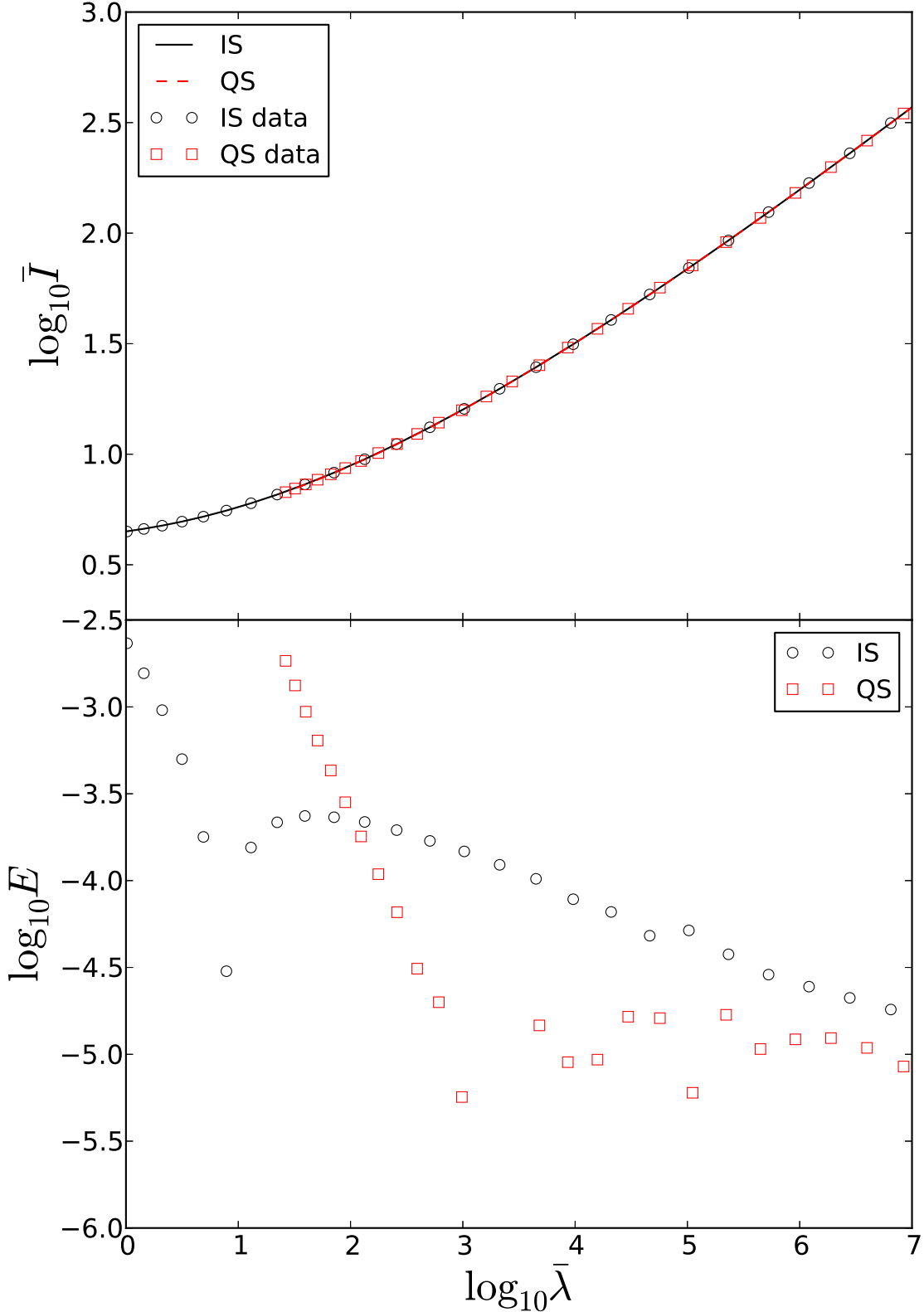}
\end{figure}

First of all, we examine the accuracy of the above
post-Minkowskian expansions of the I-Love relation for ISs and
QSs. As shown in Fig.~\ref{fig-error}, where the I-Love relation
obtained numerically for ISs and QSs are compared with the 7-term
post-Minkowskian expansions \eqref{IS-ILove} and \eqref{QS-ILove},
both of these two expansions can accurately reproduce the I-Love
relation for these two kinds of SBSs with the relative error $E
\equiv |(\bar{I})_{\rm series}/( \bar{I})_{\rm data}-1| $ being
less than $0.01$ in all cases. In fact, unless for stars close to
the maximum compactness $\cc_m$ ($\cc_m =4/9, 0.275$ for ISs and
QSs, respectively), $E$ is less than $0.001$. Furthermore, it can
be readily observed from Fig.~\ref{fig-error} that the difference
between the I-Love relations for ISs and QSs (respectively denoted
by the continuous and dashed lines) is almost indiscernible. This
observation is in good agreement with the discovery  reported in
\citep{ILoveQ_1,ILoveQ_3}, which reveals the stationarity of the
I-Love relation about the IS limit.

The similarity between  the I-Love relations for ISs and QSs  is
intriguing. In Table~\ref{compare-1} we compare the corresponding
expansion coefficients in the $\bar{I}-\cc$,
$\bar{\lambda}_\cc-\cc $ and $\bar{I}-\bar{\lambda}$ relations for
these two kinds of SBSs. Comparing \eqref{IS-I} with \eqref{QS-I},
and \eqref{IS-k2} with \eqref{QS-k2}, we can see that for these
two kinds of SBSs, the respective leading first terms in the
post-Minkowskian expansion for the $\bar{I}-\cc$ and
$\bar{\lambda}_\cc-\cc $ relations are the same, whereas as shown
in  \eqref{IS-ILove} and \eqref{QS-ILove}  the leading {\it two
orders} of the post-Minkowskian expansion for the
$\bar{I}-\bar{\lambda}$ relation are identical. Besides, barring
one exceptional case, the magnitude of the relative difference of
the corresponding coefficients in these two cases (ISs  and QSs)
grows with the order, and such differences  for the
$\bar{I}-\bar{\lambda}$ relation are usually much smaller than
their counterparts in  the $\bar{I}-\cc$ and
$\bar{\lambda}_\cc-\cc $ relations.

\begin{figure}
\centering \caption{The logarithm of the relative difference
between the scaled moment of inertia of QSs and ISs,
$\bar{I}_{QS}$ and $\bar{I}_{IS}$, is plotted against the
compactness $\cc$. The analytic results  obtained from the leading
7-term post-Minkowskian expansion given by \eqref{IS-I} and
\eqref{QS-I} (denoted by the continuous line) agree nicely with
the numerical result (denoted by circles).  The relative
difference is also analyzed in terms of individual contributions
due to the 1st, 2nd, $\ldots, 6{\rm th}$ post-Minkowskian
correction terms in \eqref{IS-I} and \eqref{QS-I}. It is clearly
shown  that the first-order post-Minkowskian correction term (the
dot-dashed line) is close to the total contribution (the
continuous line) of
 the leading six correction terms. }\label{I-C}
\includegraphics[scale=0.8]{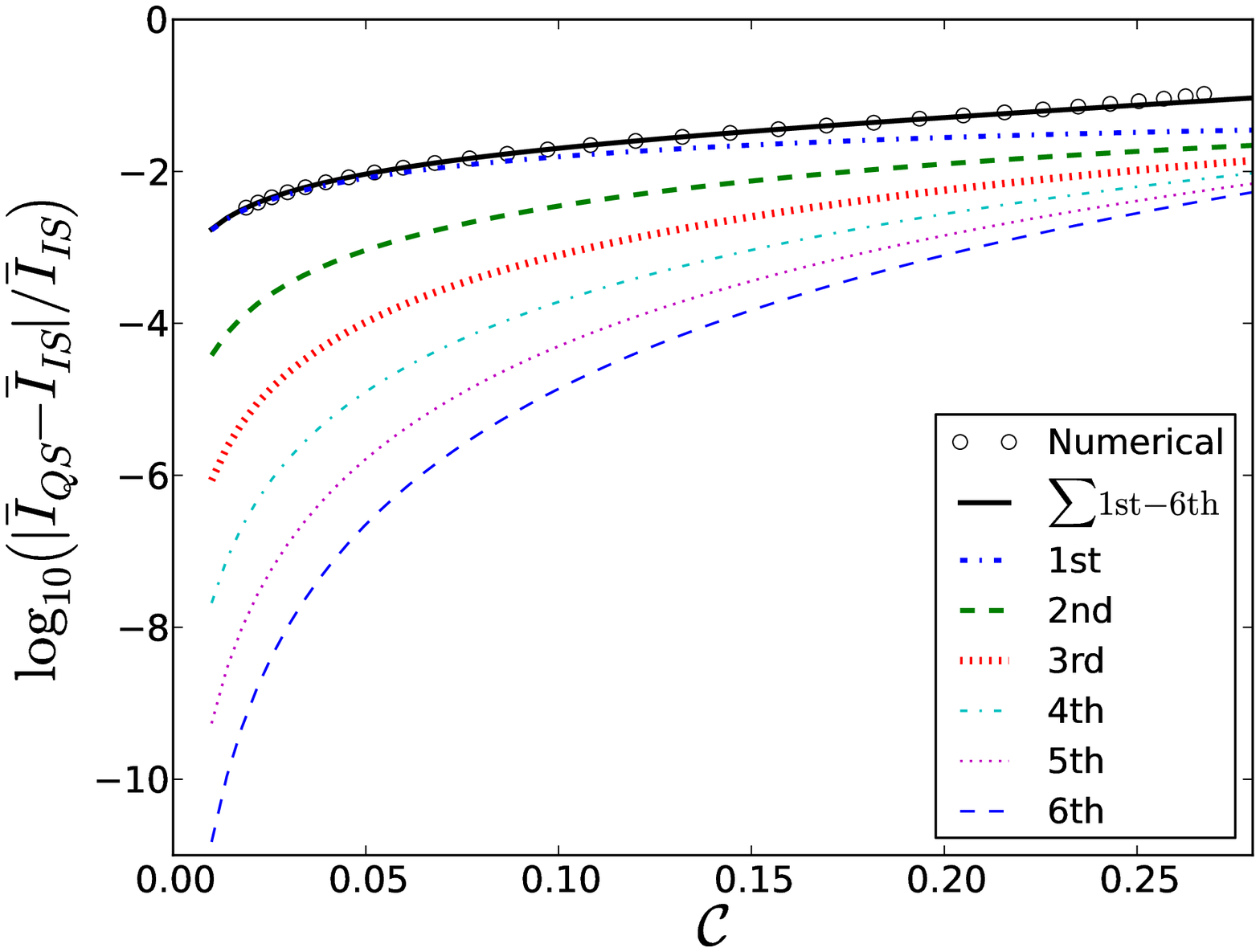}
\end{figure}

\begin{figure}
\centering \caption{The logarithm of the relative difference
between the dimensionless tidal deformability of QSs and ISs,
$\bar{\lambda}_{QS}$ and $\bar{\lambda}_{IS}$, which are obtained
from the $\bar{\lambda}_\cc-\cc$ relations (see \eqref{IS-k2} and
\eqref{QS-k2}) and \eqref{scaled_l}, is plotted against the
compactness $\cc$. The analytic results obtained from the leading
7-term post-Minkowskian expansion given by \eqref{IS-k2} and
\eqref{QS-k2} (denoted by the continuous line) agree nicely with
the numerical result (denoted by circles). The relative difference
is also analyzed in terms of individual contributions due to the
1st, 2nd, $\ldots, 6{\rm th}$ post-Minkowskian correction terms in
\eqref{IS-k2} and \eqref{QS-k2}. It is clearly shown that the
first-order post-Minkowskian correction term (the dot-dashed line)
is close to the total contribution (the continuous line) of
 the leading six correction terms.
 }\label{lambda-C}
\includegraphics[scale=0.8]{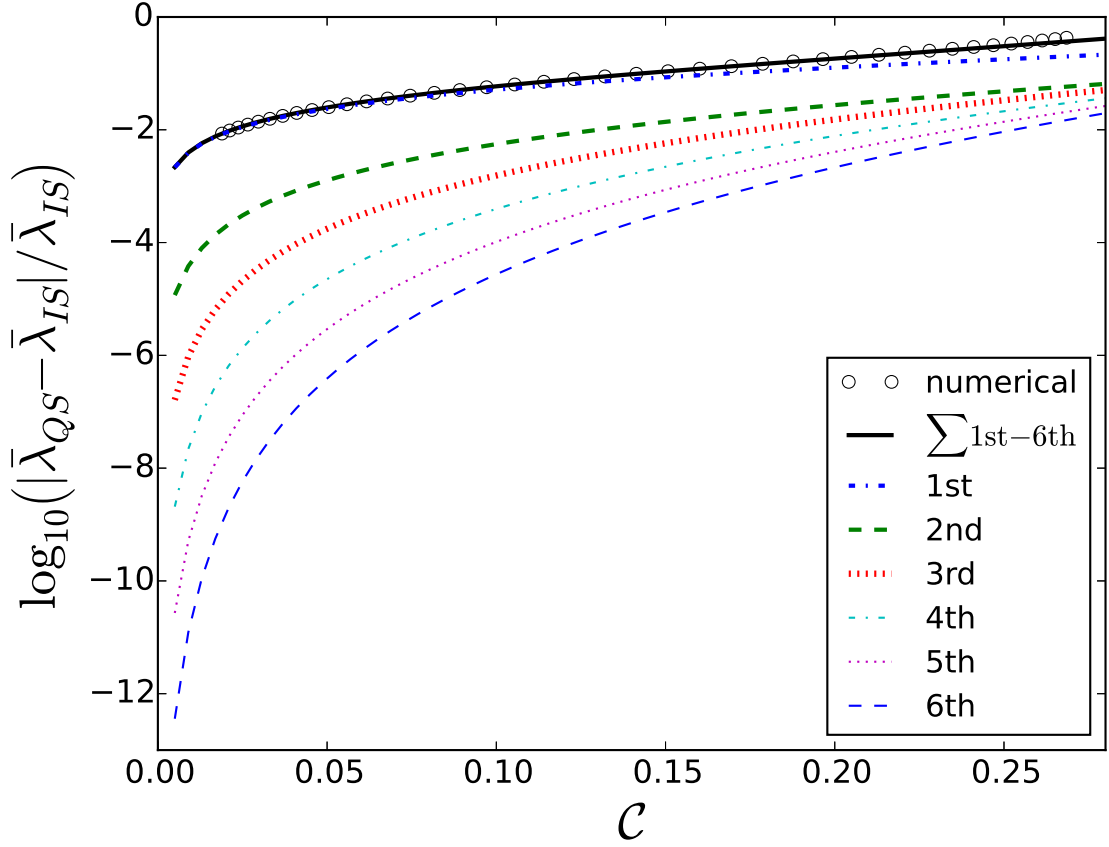}
\end{figure}

\begin{figure}
\centering \caption{The logarithm of the relative difference
between the the scaled moment of inertia of QSs and ISs,
$\bar{I}_{QS}$ and $\bar{I}_{IS}$, is plotted against
$\log_{10}\bar{\lambda}$. The analytic results  obtained from the
leading 7-term post-Minkowskian expansion given by
\eqref{IS-ILove} and \eqref{QS-ILove} (denoted by the continuous
line) agree nicely with the numerical result (denoted by circles).
 The relative difference is also analyzed in terms
of individual contributions due to the 2nd, $\ldots, 6{\rm th}$
post-Minkowskian correction terms in \eqref{IS-ILove} and
\eqref{QS-ILove}. In this case the contribution arising from the
first-order post-Minkowskian correction term vanishes and is not
shown in the figure. }\label{I-lambda}
\includegraphics[scale=0.8]{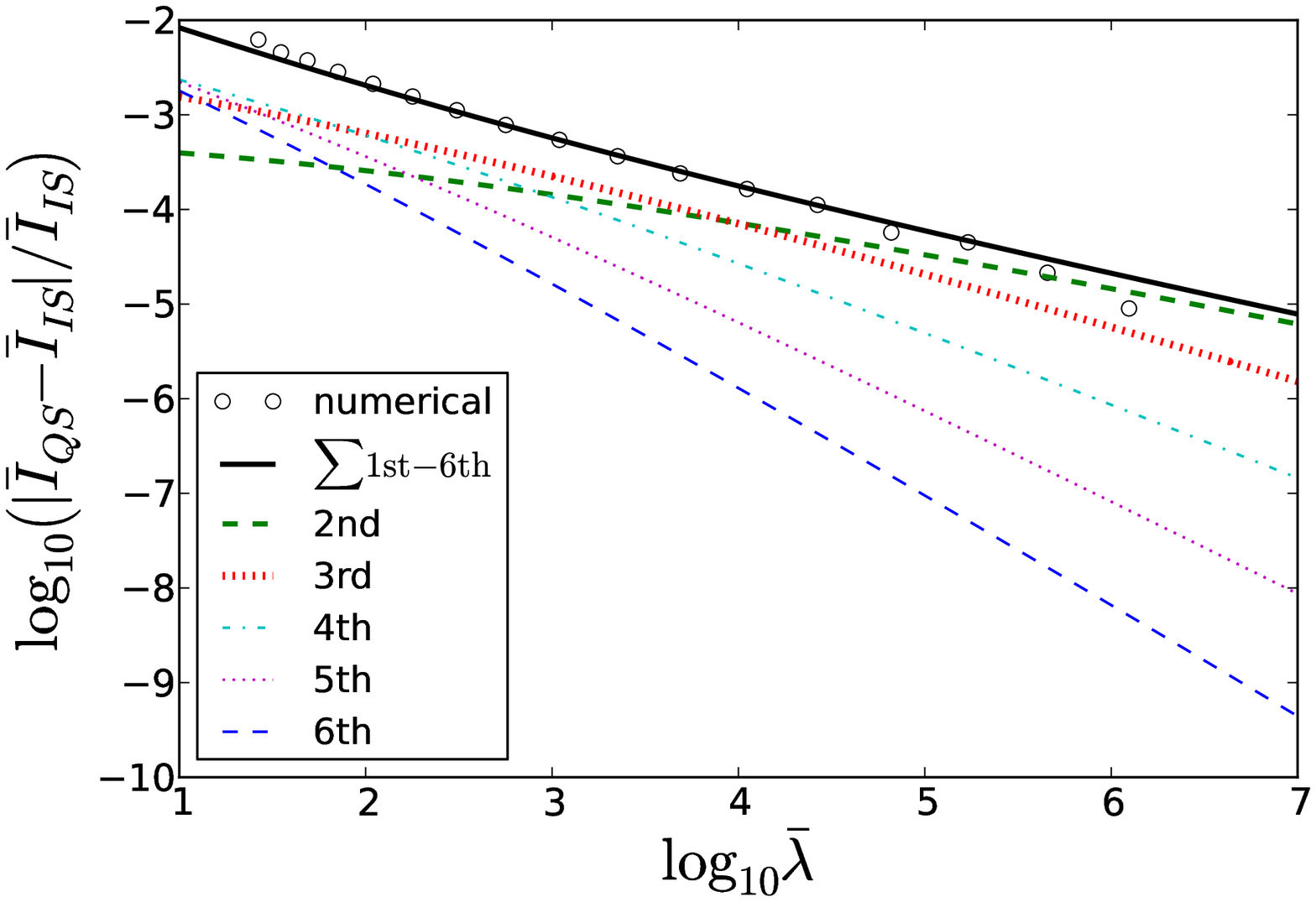}
\end{figure}

The physical consequences of the abovementioned observations are
furthered demonstrated numerically in Figs.~\ref{I-C},
\ref{lambda-C} and \ref{I-lambda}. In Fig.~\ref{I-C} the logarithm
of the relative difference between the scaled moment of inertia of
QSs and ISs, $\bar{I}_{QS}$ and $\bar{I}_{IS}$, is plotted against
the compactness $\cc$. It is clearly shown that the analytic
result (continuous line) obtained from the leading 7-term
post-Minkowskian expansion given by \eqref{IS-I} and \eqref{QS-I}
agree nicely with the numerical result (denoted by circles). As
the leading terms (referred to as the zeroth-order term hereafter)
in \eqref{IS-I} and \eqref{QS-I} cancel the contribution of each
other, the relative difference in $\bar{I}$ vanishes in the
Newtonian limit. However, the relative difference grows gradually
with $\cc$. In fact, the relative difference is larger than 0.01
for $\cc >0.1$. Such situation also prevails in
Fig.~\ref{lambda-C} where the relative difference between
$\bar{\lambda}_{QS}$  and $\bar{\lambda}_{IS}$, which are obtained
from the $\bar{\lambda}_\cc-\cc$ relations (see \eqref{IS-k2} and
\eqref{QS-k2}) and \eqref{scaled_l}, is plotted against the
compactness $\cc$. In this case the relative difference is much
larger than 0.01 for $\cc >0.1$ and grows beyond 0.1 for $\cc
>0.15$. In both Figs.~\ref{I-C} and \ref{lambda-C} we have also
decomposed the relative difference into the contributions due to
the 1st, 2nd, $\ldots, 6{\rm th}$ post-Minkowskian correction
terms in the $I-\cc$
 and $\bar{\lambda}_\cc-\cc$ relations for ISs and QSs. It is clearly shown
 that the first-order post-Minkowskian correction term (the dot-dashed line) in fact
 contributes the most to the relative difference of $\bar{I}$ and
 $\bar{\lambda}$. As a result, the first-order post-Minkowskian correction
 term in the $\bar{I}-\cc$  and $\bar{\lambda}_\cc-\cc$ relations can readily account for difference in $\bar{I}$ and
 $\bar{\lambda}$ between QSs and ISs observed previously (see,
 e.g., \citep{Lattimer:2001,Lattimer_Love}).

However, when the $\bar{I}-\cc$  and $\bar{\lambda}-\cc$ relations
are combined together to eliminate the variable $\cc$ so as to
express $\bar{I}$ directly in terms of $\bar{\lambda}$, the
EOS-dependency on the first-order post-Minkowskian correction term
of the I-Love relation surprsingly  disappears (see
\eqref{IS-ILove}, \eqref{QS-ILove} and Table~\ref{compare-1}). In
Fig.~\ref{I-lambda} we show $\log_{10}
(|\bar{I}_{QS}-\bar{I}_{IS}|/ \bar{I}_{IS})$ versus
$\log_{10}\bar{\lambda}$. Again the analytical result (the
continuous line) obtained from \eqref{IS-ILove} and
\eqref{QS-ILove} can well approximate the numerical data
(circles). In comparison with Figs.~\ref{I-C} and \ref{lambda-C},
the relative difference is now much smaller in most situations. It
is always less than 0.01 and decreases rapidly with increasing
$\bar{\lambda}$ (i.e., smaller compactness). Unless for QSs close
to the maximum compactness, the relative difference between ISs
and QSs is less than 0.001, which is ten times less than the
typical value of its counterparts shown in Figs.~\ref{I-C} and
\ref{lambda-C}. To further understand the smallness of such
difference,
in Fig.~\ref{I-lambda} we decompose $|\bar{I}_{QS}-\bar{I}_{IS}|/
\bar{I}_{IS}$  into respective contributions arising from the
leading six order post-Minkowskian expansions. Unlike the
situations shown in Figs.~\ref{I-C} and \ref{lambda-C}, where the
relative difference is dominated by the first-order expansion, in
this case the first-order post-Minkowskian correction term
 vanishes identically, which is exactly the reason why the relative
difference becomes so small. Instead, for stars with
$\log_{10}\bar{\lambda}>5$, the second-order correction term,
which is less than $10^{-4}$, contributes the most to the relative
difference. On the other hand, for more compact stars, the
contributions due to higher order post-Minkowskian corrections
overtake the second-order one. However, even for QSs close to the
maximum compactness limit, the relative difference is still
bounded by $10^{-2}$.


\begin{table}[t]
\begin{tabular}{l|lllllll}
relation&~~$n=0$~~&~~~~1~~~~&~~~~2~~~~&~~~~3~~~~&~~~~4~~~~&~~~~5~~~~&~~~~6~~~~\\
\hline
  $\bar{I}-{\cal C}$&~~~~0&~-0.20&~-0.381&~-0.635&~-1.05&~-1.769&~-3.03\\
$\bar{\lambda}_\cc-{\cal C}$ & ~~~~0 & ~~0.25 &~-0.845 & ~~22.8 &
~~12.9 &
~~13.0 &~ 14.9\\
  $\bar{I}-\bar{\lambda}$&~~~~0&~~0&~-0.00144&~-0.0174&~-0.223&~~1.90&~~2.76\\
  \hline
\end{tabular}
\caption{A comparison of the post-Minkowskian expansion of the
$\bar{I}-{\cal C}$, $\bar{\lambda}_\cc-{\cal C}$ and
$\bar{I}-\bar{\lambda}$ relations for ISs (see \eqref{IS-I},
\eqref{IS-k2} and \eqref{IS-ILove}) and QSs (see \eqref{QS-I},
\eqref{QS-k2} and \eqref{QS-ILove}). For each relation, the
relative difference between the coefficients of the $n$th term in
 the post-Minkowskian expansions for these two cases,
i.e., (the coefficient for QS case)/(the corresponding coefficient
for IS case) $-$ 1, is shown for $n=0,1,2,\cdots,6$. }
\label{compare-1}
\end{table}

In order to obtain a more thorough  understanding of these
results, we consider a general linear EOS $\rho = c_0 + c_1 p$ for
SBSs and carry out the perturbative scheme to find the formulas
for the $\bar{I}-{\cal C}$, $\bar{\lambda}_\cc-{\cal C}$ and
$\bar{I}-\bar{\lambda}$ relations. In Table~\ref{c1depend} we show
the expansion coefficients of these relations. We note that the
expansion coefficients are, as mentioned previously, independent
of $c_0$.  The general terms in $\bar{I}-{\cal C}$ and
$\bar{\lambda}_\cc-{\cal C}$ expansions are given by $c_1^m {\cal
C}^n$ ($m=0,1,\cdots$) with $n=-2,-1,\cdots$ for the former and
$n=0,1,\cdots$ for the latter. In addition, it can be seen that
\begin{eqnarray}
    \frac{1}{\bar{I}}\left(\frac{\partial \bar{I}}{\partial c_1}\right)_\cc
 &= &
    -0.05715 \cc + O(\cc^2)~,\label{eq:I-C} \\
    \frac{1}{\bar{\lambda}}\left(\frac{\partial \bar{\lambda}}{\partial c_1}\right)_\cc
    & = &
    -0.1428 \cc +O(\cc^2)~. \label{L-C}
\end{eqnarray}
Hence, both  $\bar{I}$ and $\bar{\lambda}$ are independent of the
variation in $c_1$ in the Newtonian limit where $\cc \rightarrow
0$. However, we expect that  their dependency on $c_1$ is more
obvious for relativistic stars. In fact, neither of these two
derivatives  vanishes to first-order in the compactness.

On the other hand, $\cc$ can be eliminated from the $\bar{I}-{\cal
C}$ and $\bar{\lambda}_\cc-{\cal C}$ relations to yield the
$\bar{I}-\bar{\lambda}$ relation. As shown in
Table~\ref{c1depend}, the general term of such an expansion is
$c_1^m \bar{\lambda}^{n/5}$-term, where  $m=0,1,\cdots$ and
$n=-4,-3,\cdots,2$. We also find that
\begin{eqnarray}
    \frac{1}{\bar{I}}\left(\frac{\partial \bar{I}}{\partial c_1}\right)_{\bar{\lambda}}
   &= &
    -0.002061 {\bar{\lambda}}^{-2/5} + O({\bar{\lambda}}^{-3/5})~,\nonumber \\
    &=&-0.002720 \cc^{2} + O(\cc^{3})~, \label{I-Love-c}
\end{eqnarray}
where the second line follows from the fact that ${\bar{\lambda}}=
0.5 \cc^{-5} [1+ O(\cc^{})] $ (see Table~\ref{c1depend}).
Eq.~\eqref{I-Love-c} clearly demonstrates the reason why the
I-Love relation is so insensitive to the value of $c_1$, which is
a measure of the compressibility of the stellar matter. The
logarithmic derivative of $\bar{I}$ with respect to $c_1$ at a
constant $\bar{\lambda}$, $(\partial \ln \bar{I}/\partial
c_1)_{\bar{\lambda}}$, is approximately equal to $-0.002720\cc^2$.
In comparison with the results shown in \eqref{eq:I-C} and
\eqref{L-C}, the dependency of the I-Love relation on the
parameter $c_1$ (or compressibility) is much weaker than that of
the $\bar{I}-{\cal C}$ and $\bar{\lambda}-{\cal C}$ relations. In
particular, to first-order in $\cc$, $(\partial \ln
\bar{I}/\partial c_1)_{\bar{\lambda}}$ vanishes. This clearly
explains why the I-Love relation is so insensitive to the value of
$c_1$ ($c_1=0,3$ for ISs and QSs respectively)  as observed here
and other previous publications
\citep{Yagi_14_PRD,ILoveQ_1,ILoveQ_2}.


\begin{table}[t]
\caption{The coefficients of $c_1^m {\cal C}^n$-term
($m=0,1,\cdots,6$) in the series expansions of $\bar{I}$ (with
$n=-2,-1,\cdots,4$) and $\bar{\lambda}_{\cc} $ (with
$n=1,2\cdots,6$) are shown for a SBS with a linear EOS given by
$\rho = c_0 + c_1 p$. Similarly, the coefficients of $c_1^m
\bar{\lambda}^{n/5}$-term ($m=0,1,\cdots,6$ and
$n=-4,-3,\cdots,2$) in the series expansion of $\bar{I}$
 are also tabulated.
}
\begin{tabular}{c|ccccccc}
 $\bar{I}$&$c_1^0$&$c_1^1$&$c_1^2$&$c_1^3$&$c_1^4$&$c_1^5$&$c_1^6$\\
\hline
${\cal C}^{-2}$&$4.000 \times 10^{-1}$ & 0. & 0. & 0. & 0. & 0. & 0. \\
${\cal C}^{-1}$& $3.429 \times 10^{-1}$ & $-2.286 \times 10^{-2}$
& 0. & 0. & 0. & 0. &
   0. \\
${\cal C}^{0}$& $4.038 \times 10^{-1}$ & $-4.063 \times 10^{-2}$ & $-3.556 \times 10^{-3}$ & 0. & 0. & 0. & 0. \\
${\cal C}^{1}$& $5.472 \times 10^{-1}$ & $-6.612 \times 10^{-2}$ & $-1.431 \times 10^{-2}$  & $-7.454 \times 10^{-4}$ & 0. & 0. & 0. \\
${\cal C}^{2}$& $8.035 \times 10^{-1}$ & $-1.069 \times 10^{-1}$ & $-4.219 \times 10^{-2}$ & $-4.757 \times 10^{-3}$ & $-1.852 \times 10^{-4}$ & 0. & 0. \\
${\cal C}^{3}$& 1.243 & $-1.735 \times 10^{-1}$ & $-1.105\times
10^{-1}$ &
   $-1.998 \times 10^{-2}$ & $-1.625\times 10^{-3}$ & $-5.106 \times 10^{-5}$ & 0. \\
${\cal C}^{4}$& 1.995 & $-2.824 \times 10^{-1}$ & $-2.727 \times
10^{-1}$ &
   $-6.977 \times 10^{-2}$ & $-8.833\times 10^{-3}$ & $-5.704\times 10^{-4}$ & $-1.507\times 10^{-5}$\\ \hline
$\bar{\lambda}_{\cc}$&$c_1^0$&$c_1^1$&$c_1^2$&$c_1^3$&$c_1^4$&$c_1^5$&$c_1^6$\\
\hline ${\cal C}^0$ & $7.500\times 10^{-1}$  & $0.$  & $0.$ & $0.$
& $0.$  & $0.$  & $0.$\\
${\cal C}^1$ & $-1.286$               & $1.071\times 10^{-1}$  & $0.$                   & $0.$                   & $0.$                   & $0.$                         & $0.$                         \\
${\cal C}^2$ & $4.116\times 10^{-1}$  & $-7.823\times 10^{-2}$ & $-1.259\times 10^{-2}$ & $0.$                   & $0.$                   & $0.$                         & $0.$                         \\
${\cal C}^3$ & $-4.232\times 10^{-2}$ & $-2.102\times 10^{-1}$ & $-3.112\times 10^{-2}$ & $-2.060\times 10^{-3}$ & $0.$                   & $0.$                         & $0.$                         \\
${\cal C}^4$ & $-1.907\times 10^{-1}$ & $-4.630\times 10^{-1}$ & $-8.699\times 10^{-2}$ & $-9.656\times 10^{-3}$ & $-4.432\times 10^{-4}$ & $0.$                         & $0.$                         \\
${\cal C}^5$ & $-4.933\times 10^{-1}$ & $-1.014$               & $-2.348\times 10^{-1}$ & $-3.664\times 10^{-2}$ & $-3.166\times 10^{-3}$ & $-1.139\times 10^{-4}$       & $0.$                         \\
${\cal C}^6$ & $-1.152$               & $-2.213$ & $-6.172\times
10^{-1}$ & $-1.248\times 10^{-1}$ & $-1.571\times 10^{-2}$ &
$-1.097\times 10^{-3}$       &$-3.254\times 10^{-5}$\\
\hline
$\bar{I}$&$c_1^0$&$c_1^1$&$c_1^2$&$c_1^3$&$c_1^4$&$c_1^5$&$c_1^6$\\
\hline
$\bar{\lambda}^{2/5}$&$5.278 \times 10^{-1}$  & 0. & 0. & 0. & 0. & 0. & 0.  \\
$\bar{\lambda}^{1/5}$& 1.444 & 0. & 0. & 0. & 0. & 0. &  0.\\
$\bar{\lambda}^0$& 1.583 & $-1.088 \times 10^{-3}$  & $1.088 \times 10^{-4}$  & 0. & 0. & 0. & 0.   \\
$\bar{\lambda}^{-1/5}$& $8.187 \times 10^{-1}$  & $-3.731 \times 10^{-3}$  & $-3.323 \times 10^{-4}$  & $-9.845 \times 10^{-7}$  & 0. & 0. & 0.  \\
$\bar{\lambda}^{-2/5}$& $1.528 \times 10^{-1}$  & $-4.78 \times 10^{-3}$  & $-1.665 \times 10^{-3}$  & $-1.599 \times 10^{-4}$  & $-4.691 \times 10^{-6}$  & 0. & 0.  \\
$\bar{\lambda}^{-3/5}$& $-2.686 \times 10^{-2}$  & $-2.55 \times 10^{-3}$  & $-2.565 \times 10^{-3}$  & $-5.82 \times 10^{-4}$  & $-5.259 \times 10^{-5}$  & $-1.652 \times 10^{-6}$  & 0.  \\
$\bar{\lambda}^{-4/5}$& $-2.366 \times 10^{-2}$  & $2.617 \times 10^{-8}$  & $-2.142 \times 10^{-3}$  & $-1.011 \times 10^{-3}$  & $-1.831 \times 10^{-4}$  & $-1.485 \times 10^{-5}$  & $-4.473 \times 10^{-7}$ \\
\hline
\end{tabular}
\label{c1depend}
\end{table}

\section{General case}\label{general}
The insensitivity of the I-Love relation to the parameters
characterizing the EOS is generic. In general, we consider a SBS
with EOS given by (\ref{QS_AS1}). In the following we shall show
that the I-Love relation is, to the $j$th-order post-Minkowskian
correction, independent of the parameters $c_j$ about the
incompressible limit. Without loss of generality, we assume for
the moment that for $j=1,2,3,\cdots$, $c_j=0$ if $j \ne n$, where
$n$ is a given positive integer. Therefore, all physical
quantities can be considered as functions of $c_n$ and are
expandable in terms of Taylor series in $c_n$. For example,
$\rho(x;\cc; c_n)$ and $p(x;\cc; c_n)$ are expanded as:
\begin{eqnarray}
\rho(x;\cc;
c_n)&=&\rho^{(0)}(x;\cc)+\rho^{(1)}(x;\cc)c_n+\rho^{(2)}(x;\cc)c_n^{2}+\cdots~,\label{rho-cn}\\
p(x;\cc;
c_n)&=&p^{(0)}(x;\cc)+p^{(1)}(x;\cc)c_n+p^{(2)}(x;\cc)c_n^{2}+\cdots~\label{p-cn}.
\end{eqnarray}
The physical meaning of $\rho^{(0)}(x;\cc)$ and $p^{(0)}(x;\cc)$
are obvious. They are  the density and pressure profiles,
respectively, of an IS with constant density $c_0$ and compactness
$\cc$. In the following we shall show that to first-order in $c_n$
and
 $\cc^n$, the I-Love relation of such SBSs is independent of $c_n$. In other words, to $\cc^n$, the
I-Love relation is stationary with respect to variation in $c_n$
about the incompressible limit where $c_n=0$ (see
\eqref{generalcase} for the corresponding mathematical formula).
To this end, we shall only keep terms up to first-order in $c_n$
and ignore its higher-order terms in the following calculations.

First of all, we note that to first order in $\cc$,
$p^{(0)}(x;\cc)=c_0 \cc(1-x^2)/2$, and hence it follows from the
relevant EOS that
\begin{equation}
[\rho^{(1)}(x;\cc)]_\cc=
 \alpha_n [f_n(x)-3\hat{I}_n]~,\label{rho1}
\end{equation}
where the notation $[g(\cc)]_\cc$ is introduced hereafter to
signify the leading term in the Taylor expansion of $g$ as a
function of $\cc$,
\begin{eqnarray}\label{}
\alpha_n &=& (c_0 \cc/2)^n~, \label{alpha}\\
f_n(x) &=&(1-x^2)^n~,\label{fn}\\
    \hat{I}_n &=& \int_0^1 f_n(x) x^2 dx = \frac{\sqrt{\pi} \Gamma (n+1)}{4 \Gamma
    (n+5/2)}~,\label{hatI}
\end{eqnarray}
with $\Gamma(z)$ being the standard $\Gamma$-function (see, e.g.,
\citep{MathFnHandbook}). It is worthwhile to remark that in
\eqref{rho1} there are two contributions to
$[\rho^{(1)}(x;\cc)]_\cc$, namely, $\alpha_n f_n(x)$ and
$-3\alpha_n \hat{I}_n$. While the former is the density change due
to the finite compressibility
 of the EOS, the latter is introduced to keep the mass, the radius and hence
the compactness of the star  unchanged in the process of switching
on $c_n$.

 Next, we consider the moment of inertia given by
(\ref{moi1}) and expand $\Lambda(x;\cc;c_n)$ and $ j(x;\cc;c_n)$
as follows:
\begin{eqnarray}
\Lambda(x;\cc;c_n)&=&\Lambda^{(0)}(x;\cc)+\Lambda^{(1)}(x;\cc)c_n+\Lambda^{(2)}(x;\cc)c_n^{2}+\cdots~,\\
j(x;\cc;c_n)&=&j^{(0)}(x;\cc)+j^{(1)}(x;\cc)c_n+j^{(2)}(x;\cc)c_n^{2}+\cdots~.
\end{eqnarray}
As mentioned above, $\Lambda^{(0)}(x;\cc)$ and $j^{(0)}(x;\cc)$
are the corresponding physical quantities for ISs, which are given
by:
\begin{eqnarray}
\Lambda^{(0)}(x;\cc)&=&\frac{2}{5}\left(5-3x^{2}\right)\cc
+\frac{1}{70}\left(21+126x^{2}-99x^{4}\right)\cc^{2}+\cdots~,\\
j^{(0)}(x;\cc)&=&1+\frac{3\cc(1-x^2)}{2}+3\cc^2(1-x^2)+\cdots~.
\end{eqnarray}
On the other hand, since
\begin{equation}\label{}
 \frac{ dj(x;\cc)}{dx}=-\frac{1}{2}\left[1-\left(\frac{\nu+\lambda}{2}\right)
 +\frac{1}{2}\left(\frac{\nu+\lambda}{2}\right)^2+\cdots\right]\frac{d(\nu+\lambda)}{dx}~
 \end{equation}
and $d(\nu+\lambda)/dr=8\pi(p+\rho)/(1-2m/r)$, it is
straightforward to show from the TOV equations and \eqref{moi3}
that
\begin{equation}\label{}
\left[ \frac{ dj^{(1)}(x;\cc)}{dx}\right]_\cc=\frac{3 \cc
\alpha_n}{c_0}
 \left[3\hat{I}_n-f_n(x)\right]x~,
 \end{equation}
and
 \begin{equation}
\left[\frac{d\Lambda^{(1)}}{dx}\right]_\cc =\frac{12 \cc
\alpha_n}{x^{4}c_0}\int_{0}^{x}x'^{4}\left[3\hat{I}_n-f_n(x')\right]{dx'}
 ~.
\end{equation}
At $x=1$, $\Lambda=-\Lambda'/3=2\cc^3 \bar{I}$, we have
 \begin{eqnarray}
\left[\Lambda^{(1)}\right]_\cc(x=1) &=&\frac{4 \cc
}{c_0}\int_{0}^{1}x^{4} [\rho^{(1)}(x;\cc)]_\cc
\,{dx}
 ~,\nonumber \\
&=& -\frac{6 \cc \alpha_n}{5 c_0} \left(\frac{n}{2n+5}\right)
\frac{\sqrt{\pi} \Gamma (n+1)}{ \Gamma
    (n+5/2)}~,\label{lambda-1}
\end{eqnarray}
and hence
\begin{eqnarray}
   [ \bar{I}^{(1)}]_\cc&=&\frac{2
}{c_0 \cc^2}\int_{0}^{1}x^{4} [\rho^{(1)}(x;\cc)]_\cc \,{dx}
 ~,\nonumber \\
   &=&-\frac{3  \alpha_n}{5 c_0 \cc^2} \left(\frac{n}{2n+5}\right)
\frac{\sqrt{\pi} \Gamma (n+1)}{ \Gamma
    (n+5/2)}~.\label{I-1}
\end{eqnarray}
We note that
\begin{equation}\label{I-0}
 \bar{I}^{(0)}=\frac{1}{\cc^2}\left[\frac{2}{5}+\frac{12}{35}{\cal C}+\frac{212}{525}{\cal C}^2+
\frac{632}{1155}{\cal C}^3+\frac{703744}{875875}{\cal
C}^4+\frac{251264}{202125}{\cal C}^5+
\frac{121542272}{60913125}{\cal C}^6+\cdots~\right],
\end{equation}
which is the dimensionless moment of inertia of ISs.

We now turn to the problem of tidal deformability and consider the
function  $y(r)$ governed by \eqref{ylove-1}. Similar to the
series expansion of $I$, we assume $y$ can be written as a power
series in ${c_n}$
\begin{align}
\label{yexp}
y(x;\cc;c_n)=y^{(0)}(x;\cc)+y^{(1)}(x;\cc){c_n}+y^{(2)}(x;\cc)c_n^2+y^{(3)}(x;\cc)c_n^3+\cdots
\end{align}
We substitute  \eqref{yexp} and the series expansions of other
quantities into \eqref{ylove-1}. The zeroth-order term leads to
\begin{align}
\label{ylove_0} r
\frac{dy^{(0)}(r)}{dr}+y^{(0)}(r)^2&+y^{(0)}(r)e^{\lambda^{(0)}(r)}[1+4\pi
r^2(p^{(0)}(r)-c_0)]+r^2Q^{(0)}(r)=0,
\end{align}
where
\begin{align}
Q^{(0)}(r;\cc)=&4\pi
e^{\lambda^{(0)}(r)}\left[5c_0+9p^{(0)}(r)\right]-\frac{6e^{\lambda^{(0)}(r)}}{r^2}-\left[\frac{d\nu^{(0)}(r)}{dr}\right]^2
,
\end{align}
with $\lambda^{(0)}(r)$ and  $\nu^{(0)}(r)$ being the metric
coefficients of ISs. $y^{(0)}$ is nothing but the solution of $y$
for ISs and in Section~\ref{Love} we have already shown that
$y^{(0)}(x,\cc)=\sum_{i=0}^\infty y^{(0)}_i(x)\cc^i$, where
\begin{eqnarray}
\label{firsty} x \frac{dy^{(0)}_0(x)}{dx}+y^{(0)}_0(x)^2+y^{(0)}_0(x)-6&=&0,\\
x\frac{dy^{(0)}_1(x)}{dx}+[1+2y^{(0)}_0(x)]y^{(0)}_1(x)-x^2y^{(0)}_0(x)+3
x^2&=&0,
\end{eqnarray}
and higher-order equations can be similarly obtained. The leading
coefficients are $y^{(0)}_0(x)=2$, $y^{(0)}_1(x)=-x^2/7$, and
higher-order solution can be obtained recursively.

Now we pay special attention to the leading order term in
$y^{(1)}$ (i.e., $[y^{(1)}]_\cc$), which satisfies
\begin{equation}\label{yexp_1}
r \left[\frac{dy^{(1)}}{dr}\right]_\cc+5 \left[y^{(1)}
\right]_\cc+2\left [  \left\{e^{\lambda}+4\pi
 r^2 e^{\lambda}(p-\rho) \right\}^{(1)}\right]_\cc+r^2\left [
Q^{(1)}\right]_\cc=0.
\end{equation}
Taking into account the fact that $c_s^2 =1/ (n c_n p^{n-1})$, it
can be shown that the third and the fourth terms in the LHS of the
above equation are first- and zero-order in $\cc$, respectively.
To leading order in $\cc$, the former is then negligible, while
the latter is given by:
\begin{equation}\label{}
    r^2\left [ Q^{(1)}\right]_\cc = \frac{6n
    \alpha_n}{c_0}x^2 (1-x^2)^{n-1}~.
\end{equation}
From this result and \eqref{yexp_1} we show that
\begin{eqnarray}
  \left[\hat{y}^{(1)}\right]_\cc &\equiv&
  \left[y^{(1)}(x=1)\right]_\cc=-\frac{15 \alpha_n
    }{  c_0} \int_0^1 x^4 f_n(x) dx ~.\label{y1}
\end{eqnarray}


After solving \eqref{ylove-1} by the method mentioned above, $k_2$
can be obtained from $y_R$ by expanding \eqref{k2eq} into power
series of $\cc$ \citep{Damour:09p084035,Lattimer_Love}:
\begin{equation}\label{k2_expand}
 k_2=\frac{2-{y_R}}{2 ({y_R}+3)} +\frac{5 {\cal C} \left({y_R}^2+2 {y_R}-6\right)}{2
   ({y_R}+3)^2}-\frac{5
{\cal C}^2 \left(11 {y_R}^3+66 {y_R}^2+52 {y_R}-204\right)}{14
({y_R}+3)^3}+\cdots
\end{equation}
By virtue of \eqref{y1} and \eqref{k2_expand}, the leading
correction term in $k_2^{(1)}$ is given by
\begin{equation}\label{}
  \left[k_2^{(1)}\right]_\cc  = -\frac{3
    }{ 4 } \left(\frac{5\left[\hat{y}^{(1)}\right]_\cc }{6}+ \frac{15 \alpha_n \hat{I}_n}{2
    c_0}\right)~,
\end{equation}
where the second term inside the bracket on the RHS of the
equation accounts for the correction in the discontinuity of $y$
across the stellar surface due to the presence of $\rho^{(1)}$. We
can also rewrite the above equation as follows:
 \begin{eqnarray}
\left[k_2^{(1)}\right]_\cc  &=&\frac{75 }{8
c_0}\int_{0}^{1}x^{4}[\rho^{(1)}(x;\cc]_\cc \,{dx}~,
 ~\nonumber \\
&=& -\frac{45 \alpha_n}{16 c_0} \left(\frac{n}{2n+5}\right)
\frac{\sqrt{\pi} \Gamma (n+1)}{ \Gamma
    (n+5/2)}~,\label{k2-1}
\end{eqnarray}
where the similarity between $[\bar{\Lambda}^{(1)}]_\cc(x=1)$ (see
\eqref{lambda-1}) and $[k_2^{(1)}]_\cc$ is explicitly shown. Since
$\bar{\lambda} \equiv 2k_2/(3\cc^5)$, we have
\begin{eqnarray}
\bar{\lambda}^{(0)}&=&\frac{1}{2 {\cal C}^5}-\frac{20}{7 {\cal
C}^4}+\frac{2515}{441
   {\cal C}^3}-\frac{51550}{11319 {\cal C}^2}+\frac{3347350}{3090087
   {\cal C}}+\frac{4326424}{64891827}+ \frac{368458100}{9438155727}{\cal
   C}+\cdots,\label{lambdaC-0}\\
\left[\bar{\lambda}^{(1)}\right]_\cc &=& \frac{25 }{4 c_0
\cc^5}\int_{0}^{1}x^{4}[\rho^{(1)}(x;\cc]_\cc \,{dx}~,
 ~\nonumber \\
&=&
-\frac{15}{8\cc^5}\left(\frac{\alpha_n}{c_0}\right)\left(\frac{n}{2n+5}\right)
\frac{\sqrt{\pi} \Gamma (n+1)}{\Gamma
    (n+5/2)}~.\label{lambdaC-1}
\end{eqnarray}

By elementary theory of calculus, it is easy to show that
\begin{equation}\label{partial}
  \left(\frac{\partial \bar{I}}{\partial c_n}\right)_{\bar{\lambda}}=\left(\frac{\partial \bar{I}}{\partial
  c_n}\right)_\cc -
  \left(\frac{\partial \bar{\lambda}}{\partial c_n}\right)_\cc
  {\displaystyle
  \frac{\left(\frac{\partial \bar{I}}{\partial \cc}\right)_{c_n}}{\left(\frac{\partial \bar{\lambda} }{\partial
  \cc}\right)_{c_n}}}~.
\end{equation}
Under the approximations
$\bar{I}=\bar{I}^{(0)}+[\bar{I}^{(1)}]_\cc$ and
$\bar{\lambda}=\bar{\lambda}^{(0)}+[\bar{\lambda}^{(1)}]_\cc$, it
is straightforward to show from \eqref{partial}, the explicit
expressions of $\bar{I}^{(0)}$, $[\bar{I}^{(1)}]_\cc$,
$\bar{\lambda}^{(0)}$ and $[\bar{\lambda}^{(1)}]_\cc$ obtained
above that to order of $c_0^{n-1} \cc^{n}$,
\begin{equation}\label{generalcase}
  \frac{1}{\bar{I}}\left[\left(\frac{\partial \bar{I}}{\partial
  c_n}\right)_{\bar{\lambda}}\right]_{c_n =0}=0.
\end{equation}
In order words, to order $\cc^{n}$, the I-Love relation is
stationary with respect to changes in $c_n$ about the
incompressible limit where $c_n =0$. In fact, Eq.~\eqref{I-Love-c}
is merely a special case of \eqref{generalcase} with $n=1$.

\section{Physical interpretation}\label{physics}
In this section we discuss the physical mechanism underlying
\eqref{generalcase} and hence the stationarity of the I-Love
relation. First of all, we note that both $[\bar{I}^{(1)}]_\cc$
and $[\bar{\lambda}^{(1)}]_\cc$ are proportional to the integral
$\int_{0}^{1}x^{4}[\rho^{(1)}(x;\cc)]_\cc \,{dx}$, which is
nothing but the change of the Newtonian moment of inertia of the
star upon the introduction of finite compressibility to the EOS.
This once again highlights the Newtonian nature of the observed
I-Love universality. Secondly, we can also see that to leading
order in compactness,
\begin{eqnarray}
\frac{[\bar{I}^{(1)}]_\cc}
{\bar{I}^{(0)}(\cc)}&=&\frac{5}{{c_0}}\int_{0}^{1}x^{4}[\rho^{(1)}(x;\cc)]_\cc
\,{dx}~, \label{ratio-1}\\
 \frac{[\bar{\lambda}^{(1)}]_\cc}
{\bar{\lambda}^{(0)}(\cc)}&=&\frac{25}{2{c_0}}\int_{0}^{1}x^{4}[\rho^{(1)}(x;\cc)]_\cc
\,{dx}~, \label{ratio-2}
\end{eqnarray}
where ${\bar I}^{(0)}(\cc)$ and  $\bar{\lambda}^{(0)}(\cc)$ are
the scaled moment of inertia and the tidal deformability of an IS
with the same compactness $\cc$. Therefore, as noted in some
previous studies (see, e.g., \citep{Lattimer_Love}), for SBSs
(such as QSs) both $\bar{I}$ and $\bar{\lambda}$ demonstrate an
observable EOS-dependency, which is characterized by $c_n$ in the
present situation. However, it is interesting to find that
\begin{equation}\label{ratio-3}
\frac{\left\{[\bar{I}^{(1)}]_\cc /\bar{I}^{(0)}\right\}}
{\left\{[\bar{\lambda}^{(1)}]_\cc /\bar{\lambda}^{(0)}\right\}} =
\frac{2}{5}~,
\end{equation}
which is actually independent of the functional form of
$[\rho^{(1)}(x;\cc)]_\cc$.

On the other hand, if the compactness  of an IS varies from $\cc$
to  $\cc+\delta \cc$, where $|\delta \cc| \ll 1$, the associated
change in $\bar{I}$ and $\bar{\lambda}$, namely $\delta\bar{I}$
and $\delta\bar{\lambda}$, are respectively given by (to leading
orders in $\delta \cc$ and $\cc$)
\begin{eqnarray}
\frac{\delta \bar{I}} {\bar{I}^{(0)}}&=&-2\left(\frac{\delta
\cc}{\cc}\right)~, \label{ratio-4}\\
 \frac{\delta
\lambda} {\bar{\lambda}^{(0)}}&=&-5\left(\frac{\delta
\cc}{\cc}\right)~, \label{ratio-5}
\end{eqnarray}
by virtue of the fact that $\bar{I}^{(0)} \propto \cc^{-2}$ and
$\bar{\lambda}^{(0)} \propto \cc^{-5}$ in the Newtonian limit. As
a result, we have
\begin{equation}\label{ratio-6}
\frac{\left\{\delta \bar{I} /\bar{I}^{(0)}\right\}} {\left\{\delta
\bar{\lambda}  /\bar{\lambda}^{(0)}\right\}} = \frac{2}{5}~.
\end{equation}
 Then it follows directly from \eqref{ratio-1} -
\eqref{ratio-6} that the values of $\bar{I}$ and $\bar{\lambda}$
of a SBS with compactness $\cc$ and finite compressibility can be
obtained from the their counterparts of an IS with a modified
compactness $\cc+\delta \cc $   if
\begin{equation}\label{ratio-7}
\left(\frac{\delta \cc}{\cc}\right)
=-\frac{5c_n}{2{c_0}}\int_{0}^{1}x^{4}[\rho^{(1)}(x;\cc)]_\cc
\,{dx}~.
\end{equation}
 In
other words, for an IS the respective effects of  (i) introducing
finite compressibility,  and (ii) properly adjusting the stellar
compactness on $\bar{I}$ (or $\bar{\lambda}$) are
indistinguishable in the low compactness limit provided that
\eqref{ratio-7} holds. As a result, when $\bar{I}$ and
$\bar{\lambda}$ of SBSs with finite compressibility are directly
linked to each other by eliminating $\cc$ in the $\bar{I} - \cc$
and $\bar{\lambda} -\cc$ relations, their mutual dependency is
almost identical to that of ISs to leading order in compactness.
This provides a physically transparent interpretation of the
I-Love universality.


A word of caution about the validity of the I-Love universality is
in order. While Eqs. \eqref{ratio-1} and \eqref{ratio-2} follow
closely from the physical content of moment of inertia and tidal
deformation, Eqs. \eqref{ratio-4} and \eqref{ratio-5} are actually
the consequence of the judicious definitions (or the proper
normalization)  of $\bar{I}$ and $\bar{\lambda}$, which lead to
the dependency $\bar{I}^{(0)} \propto \cc^{-2}$ and
$\bar{\lambda}^{(0)} \propto \cc^{-5}$ in the Newtonian limit. For
example, if the tidal Love number $k_2 = 3 \cc^5 \bar{\lambda}/2$
is considered instead of the dimensionless tidal deformability
$\bar{\lambda}$ in the I-Love relation, the universality will no
longer prevail.
\section{Conclusion and discussion}\label{Conclusion}
In the present paper we study the physical mechanism underlying
the universality demonstrated in the I-Love relation. Motivated by
the recent numerical observation  that the I-Love relation of
realistic compact stars (including NSs and QSs) can be well
approximated by that of ISs \citep{ILoveQ_1,ILoveQ_3}, we carry
out an in-depth
 investigation on  this relation for QSs and other SBSs as well. To set the stage for such analysis,
 we establish a systematic perturbative scheme to evaluate the moment of inertia and the Love number (tidal deformability)
 of SBSs. Both $\bar{I}$ and $\bar{\lambda}$ are written in terms of power series of $\cc$.
 It can be seen from such expansions how these two physical
 quantities depend on the EOS and compactness of a SBS star. In general,
 each of the  parameters $c_n$
 in EOS \eqref{QS_AS1} can lead to a fractional deviation  in $\bar{I}$ (or $\bar{\lambda}$)
 from its IS counterpart, which is to leading order given by $c_n\cc^n$ and becomes noticeable (of order $10\%$) for
 relativistic stars (see Figs.~\ref{I-C} and \ref{lambda-C}).
However, we further    show analytically that, due to a
cancellation between the corresponding terms in the $\bar{I} -
\cc$ and $\bar{\lambda} - \cc$ relations,  the I-Love relation is,
to order $\cc^n$ ($n=1,2,3\cdots$), stationary around the
incompressible limit upon variation of the parameter $c_n$.
The universality of the I-Love relation is hence attributable to
such stationarity. Therefore, the perturbative scheme established
here indeed provides an independent corroboration of the numerical
results obtained in Refs.~\citep{ILoveQ_1,ILoveQ_3}.

As the parameter $c_n$ is essentially a measure of the
compressibility of the stellar matter, it is clearly shown here
that high stiffness of EOS is crucial to the universality of the
relation. It is worth of remark that \citet{whyI} have suggested
the validity of the elliptical isodensity approximation as a key
factor affecting the universal I-Love relation. In particular,
they found numerically that the variation of the eccentricity of
an isodensity surface inside a slowly rotating Newtonian star is
small for stiff polytropic stars and QSs and becomes exact in the
incompressible limit \citep{whyI}. Hence, their empirical finding
is actually in agreement with the analytical study reported
here.

In the present paper we also pinpoint the physical origin of the
universality and the stationary for the I-Love relation of SBSs.
We show that, as far as $\bar{I}$ and $\bar{\lambda}$ are
concerned, the effects due to finite compressibility, which is
rooted in the EOS, and a proper renormalization of the stellar
compactness (see \eqref{ratio-7}) are equivalent to leading order
in compactness. Such equivalence leads to the observed
universality and stationarity, and is the joint consequence of (i)
the similarity in the responses of $\bar{I}$ and $\bar{\lambda}$
to variations in the EOS (see \eqref{ratio-1} and
\eqref{ratio-2}), and (ii) the judicious normalization of
$\bar{I}$ and $\bar{\lambda}$. While the former is attributable to
the nature of the two relevant physical quantities,  the latter is
indeed a clever choice engineered for the I-Love universality to
hold. We expect that our analysis established here could be
generalize  to study other systems that demonstrate similar
universality (see, e.g.,
\citep{Maselli:2013,Pappas_14_prl,Chak_14_prl,Haskell_14_mnras,Yagi_14_PRD,Yagi_hair_GR,Stein_hair_apj,ILoveQ_2}).

On the other hand, we have to stress that the stationarity of the
I-Love relation about the incompressible limit holds only up to
order $\cc^n$ ($n=1,2,3\cdots$) for variations in the parameter
$c_n$. Consequently, we expect that the accuracy of such a
``universal'' relation worsens for stars with large compactness.
This point is verified in Fig.~\ref{I-lambda} of the present paper
for QSs and similar conclusion has also been obtained previously
for realistic NSs \citep{ILoveQ_1,ILoveQ_3}. However, as the
compactness of stable NSs and QSs is usually less than 0.3, the
I-Love relation of realistic compact stars does not deviate much
from its incompressible counterpart even in the extreme
relativistic limit.

As a byproduct of the present paper, we have found analytically
the universal structure of QSs and other SBSs as well, including
the density profile, the mass-radius relation, the moment of
inertia and the tidal Love number, by employing the compactness as
an expansion parameter. Each of these analytic expressions for QSs
is expected to be of interest to astrophysicists (see, e.g.,
\citep{Lattimer:2001,Lattimer:2005p7082,MomentofInertia,Lattimer_Love}).

Throughout the present paper we have focused our attention on the
I-Love relation of SBSs, whose energy density is non-zero at the
stellar surface, because of their proximity to ISs.  We show  that
 the I-Love relation of SBSs is still close to that of ISs when relativistic effects are included.
 We also provide analytic methods to evaluate quantitatively
 such effects with post-Minkowsian expansion. Whether and how our analysis established here could be generalized
to realistic NSs with vanishing surface density is a challenging
issue that is beyond the scope of the present paper. We hope that
our work reported here could trigger more other investigations
along  a similar direction. On the other hand, we also note that
\citet{ILoveQ_1} have already used the GTM, whose density profile
is given by \eqref{GTM}, to study the I-Love relation in the
Newtonian case. They showed that the I-Love relation remains
almost unchanged (with relative error less than 0.001) as the
parameter $\delta$ in \eqref{GTM} goes from 0 to 1 (see Fig.~5 in
their paper). In particular, the I-Love relation are exactly
stationary with respect to changes in $\delta$ about the two
points $\delta=0$
 and $\delta=1$. While the former (i.e., $\delta=0$) corresponds to  ISs, the
 latter (i.e., $\delta=1$)
 actually reduces to the Tolman VII model, which has been proposed
 by \citet{Lattimer:2001} as an approximately universal  density profile for
 realistic NSs. Moreover, for $0<\delta<1$, the GTM
 could also nicely reproduce the leading behavior of the density profile of SBSs as given by
 \eqref{QS_AS9}. Inasmuch as the I-Love relation is considered in the Newtonian limit,  ISs, SBSs and realistic NSs
 are deemed equivalent to each other. We therefore hold a positive view on
 the possibility of generalizing our method to realistic NSs.

Lastly,  in the ``I-Love-Q universal relations" discovered in
Refs.~\citep{Yagi:2013long,Yagi:2013} the spin-induced quadrupole
moment $Q$ (with suitable normalization) of compact stars  is also
a member of the trio which display universal behavior. However, in
order to performed a detailed analysis on the physical nature of
the I-Love relation without further ado, we have not addressed the
issue of the spin-induced quadrupole  in the present paper.
Nevertheless, we expect that our method can be generalized to
handle the case of the spin-induced quadrupole without much
difficulty.  As a matter of fact, in the Newtonian limit, the
spin-induced quadrupole moment $Q$ (with suitable normalization)
is directly proportional the tidal quadrupole moment, which is
measured by the tidal Love number, whose universal behavior is
well studied here. Therefore, we expect that the universal
behavior of the spin-induced quadrupole can be explained in a  way
similar to the case of the other two members of the trio. The work
in such an direction is underway and relevant details will be
reported elsewhere in due course.

\begin{acknowledgments}
 We thank L.M.~Lin,  Y.H.~Sham,  H.K.~Lau and J.~Wu for helpful discussions.
 \end{acknowledgments}

\appendix
\section{Perturbative expansion for Quark Stars} To illustrate and gauge the accuracy of the
perturbative expansion developed in the present paper,  we apply
the method to study the stellar structure of QSs obeying the
simple MIT bag model \eqref{QMEOS}.
For such a linear EOS, it is readily
 shown that  under the  transformations $\bar{p}=p/B$,
 $\bar{\rho}=\rho/B$, $\bar{r}=B^{1/2}r$, $\bar{R}=B^{1/2}R$  and $\bar{M}=B^{1/2}M$
 the TOV equations are independent of $B$ and hence acquire a
scale-invariant form.

 The following shows the leading expansion of
${\bar \rho}$, ${\bar p}$ and $e^{\nu}$:
\begin{eqnarray}
{\bar \rho}(x)&=&4+6\left(1-x^{2}\right)\cc
+\frac{12}{5}\left(1-x^{2}\right)\left(8-3x^{2}\right)\cc^{2}+\cdots~~,\label{QS_AS4}
\\
{\bar p}(x)&=&2\left(1-x^{2}\right)\cc
+\frac{4}{5}\left(1-x^{2}\right)\left(8-3x^{2}\right)\cc^{2}+\cdots~~,\label{QS_AS5}\\
e^{\nu}(x)&=&1-\left(3-x^{2}\right)\cc
+\frac{3}{10}\left(1-x^{2}\right)^{2}\cc^{2}+\cdots~.\label{QS_AS6}
\end{eqnarray}
 For reference, analytic expressions of ${\bar \rho}_{n}(x) \equiv { \rho}_{n}(x)/B$ and $(e^{\nu})_{n}(x)$
for $0 \le n \le 6$ are given in Tables~\ref{TSS2} and \ref{TSS3},
respectively.   We have numerically verified that for stable QSs
(i.e., $\cc <0.275$), all of these series tend to their exact
values as the order of expansion goes to infinity. In general, the
percentage error of an expansion with a fixed order grows towards
the stellar center with strong gravity. Besides, we also find that
the $m$-th order diagonal Pad\'{e} approximant about the point
$\cc=0$ ($m=1,2,\cdots$)
for $\bar{\rho}$ and $e^{\nu}$ constructed from their respective
$2m$-th order partial sums (see, e.g., \cite{Baker} for the theory
and the construction of Pad\'{e} approximants) can significantly
improve their rate of convergence.

\begin{table}[htp]
\caption{The coefficient of the $x^j$-term in the polynomials
$\bar{\rho}_n \equiv {\rho}_n/B$ for $0 \le n \le 6$. There is no
odd power term and the coefficient vanishes if $j >2n$.
}\label{TSS2} \centering
\begin{tabular}{c|cccccccc}
&$x^0$&$x^2$&$x^4$&$x^6$&$x^8$&$x^{10}$&$x^{12}$\\
\hline
\\
$\bar{\rho}_0$&$4$&&&&&&\\
\\
$\bar{\rho}_1$&$6$&$-6$&&&&&\\
\\
$\bar{\rho}_2$&$\frac{96}{5}$&$\frac{-132}{5}$&$\frac{36}{5}$&&&&\\
\\
$\bar{\rho}_3$&$\frac{10863}{175}$&$\frac{-18636}{175}$&$\frac{1269}{25}$&$\frac{-222}{35}$&&&\\
\\
$\bar{\rho}_4$&$\frac{7047}{35}$&$\frac{-71331}{175}$&$\frac{1845}{7}$&$\frac{-10797}{175}$&$\frac{768}{175}$&&\\
\\
$\bar{\rho}_5$&$\frac{882017127}{1347500}$&$\frac{-508989846}{336875}$&$\frac{1504989}{1250}$&$\frac{-2449284}{6125}$&$\frac{27333}{500}$&$\frac{-23256}{9625}$&\\
\\
$\bar{\rho}_6$&$\frac{186826212171}{87587500}$&$\frac{-480394955907}{87587500}$&$\frac{2463177573}{481250}$&$\frac{-132119397}{61250}$&$\frac{52200429}{122500}$&$\frac{-1412469}{38500}$&$\frac{912234}{875875}$\\
\\
\hline

\end{tabular}
\end{table}
\begin{table}[h]
\caption{The coefficient of the $x^j$-term in the polynomials
$(e^{\nu})_{n}$
  for $0 \le
n \le 6$. There is no odd power term and the coefficient vanishes
if $j >2n$. }\label{TSS3} \centering
\begin{tabular}{c|cccccccc}
&$~~~~~~~~x^0~~~~~~~$&$~~~~~~~x^2~~~~~~~$&$~~~~~~x^4~~~~~~$&$~~~~~~x^6~~~~~~$&$~~~~~x^8~~~~~$&$~~~~~x^{10}~~~~~$&$~~~~~x^{12}~~~~~$\\
\hline
\\
$(e^{\nu})_0$&$1$&&&&&&\\
\\
$(e^{\nu})_1$&$-3$&$1$&&&&&\\
\\
$(e^{\nu})_2$&$\frac{3}{10}$&$\frac{-3}{5}$&$\frac{3}{10}$&&&&\\
\\
$(e^{\nu})_3$&$\frac{27}{175}$&$\frac{-123}{350}$&$\frac{6}{25}$&$\frac{-3}{70}$&&&\\
\\
$(e^{\nu})_4$&$\frac{-441}{1400}$&$\frac{1}{70}$&$\frac{579}{700}$&$\frac{-183}{350}$&$\frac{-1}{40}$&&\\
\\
$(e^{\nu})_5$&$\frac{-2426307}{1347500}$&$\frac{3361973}{2695000}$&$\frac{23007}{8750}$&$\frac{-43917}{24500}$&$\frac{-29}{100}$&$\frac{69}{11000}$&\\
\\
$(e^{\nu})_6$&$\frac{-2343752949}{350350000}$&$\frac{86948269}{15925000}$&$\frac{32187921}{3850000}$&$\frac{-774423}{122500}$&$\frac{-11639}{14000}$&$\frac{279}{11000}$&$\frac{-5601}{2002000}$\\
\\
\hline

\end{tabular}

\end{table}

On the other hand, we note that for a given order of
post-Minkowkian expansion, the fractional error of  (\ref{QS_AS4})
is usually several times  larger than that of (\ref{QS_AS6}).
In fact, the fractional error of the former  is almost ten times
of that of the latter (see Ref.~\citep{atma} for details). Hence,
it is desirable to find a way to express $\bar{\rho}$ in terms of
$e^{\nu}$, which has a series expansion with higher precision for
the same order of expansion. To this end, we find from the TOV
equations    an exact expression for ${\bar \rho}$,
\begin{equation}\label{QS_AS9}
{\bar \rho}=3(1-2\cc)^{2}e^{-2\nu}+1 ~,
\end{equation}
where the boundary conditions $e^{\nu}(x=1)=1-2\cc$ and ${\bar
\rho}(x=1)=4$ have been used to fix some constants. Therefore,
${\bar \rho}$ can be found directly from $e^{\nu}$, i.e.,
(\ref{QS_AS6}).
Even at the center of the star, the percentage error in ${\bar
\rho}$ obtained from the sixth-order post-Minkowkian expansion of
$e^{\nu}$ and \eqref{QS_AS9}
is less than  $5 \%$. If, instead, a 3-3 diagonal Pad\'{e}
approximant for $e^{\nu}$ is constructed from its sixth-order
post-Minkowkian expansion and \eqref{QS_AS9} is used in tandem to
find ${\bar \rho}$, the  percentage error of ${\bar \rho}$ is much
smaller than $1\%$ throughout the whole star.





\begin{figure}
\centering \caption{$\bar{R}$ is plotted against $\bar{M}$ by
combining \eqref{A-MR1}, \eqref{A-MR2} and the sixth-order
post-Minkowsian expansion of $\Theta$ (see \eqref{theta}). The
perturbative expansion (the continuous solid line) can accurately
reproduce the exact numerical result (the dots) for $\bar{M}$ less
than the stability limit beyond which QSs become
unstable.}\label{MR-curve}
\includegraphics[scale=0.5]{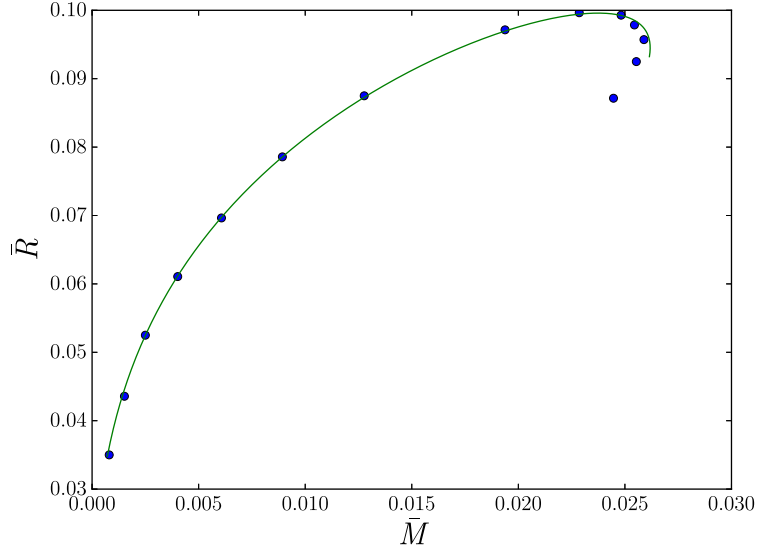}
\end{figure}

Based on  stellar profile obtained above, the mass $M$ and the
radius $R$ of QSs can be readily found as follows:
\begin{eqnarray}
R&=&\Theta\left(\frac{3\cc}{16\pi B}\right)^{1/2}~,\label{A-MR1}\\
M&=&\Theta\left(\frac{3\cc^3}{16\pi B}\right)^{1/2}~,\label{A-MR2}
\end{eqnarray}
where
\begin{eqnarray}
\Theta&=&[16\pi/(3\hat{{\cal I}})]^{1/2}\nonumber \\
&=&1-\frac{3}{10}\cc-\frac{939}{1400}\cc^{2}
-\frac{21977}{14000}\cc^{3}-\frac{32997091}{8624000}\cc^{4}-
\frac{54288534499}{5605600000}\cc^{5}-\frac{1811954651667}{71344000000}\cc^{6}+\cdots~.\label{theta}
\end{eqnarray}
Figure~\ref{MR-curve} shows $\bar{R}$ as a function $\bar{M}$,
which is obtained by combining \eqref{A-MR1}, \eqref{A-MR2} and
\eqref{theta}. It is clearly shown there that the sixth-order
post-Minkowsian expansion of $\Theta$ (the continuous solid line)
can accurately reproduce the exact numerical result (the dots) as
long as $\bar{M}$ is less than its maximum value beyond which QSs
become unstable.

\begin{figure}
\centering \caption{The values of the normalized moment of inertia
$a$ obtained from the sixth-order post-Minkowsian expansion (the
continuous solid line) and numerical integration (the dots),
respectively, are plotted against compactness $\cc$ and compared
with each other. For reference, we also include in the figure the
value of $a$ obtained from the empirical formula \eqref{empirical}
(the dashed straight line). }\label{ac-curve}
\includegraphics[scale=0.5]{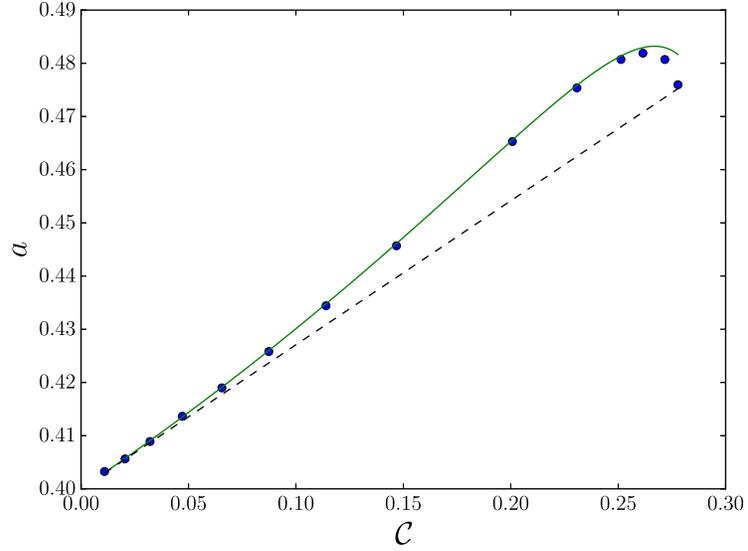}
\end{figure}

Similarly, we can also find the normalized moment of inertia $a$:
\begin{eqnarray}
a &=&
\frac{2}{5}+\frac{48}{175}\cc+\frac{656}{2625}\cc^2+\frac{40408}{202125}\cc^3
-\frac{883424}{21896875}\cc^4-\frac{24137984}{25265625}\cc^5-\frac{339641200144}{83755546875}\cc^6+\cdots~.
\label{Amoi7}
\end{eqnarray}
Figure \ref{ac-curve} compares the values of the normalized moment
of inertia $a$ obtained from the sixth-order post-Minkowsian
expansion (the continuous solid line) and numerical integration
(the dots), respectively. We see that the agreement between the
perturbative and numerical results is almost perfect unless the
compactness is close to the stability limit $\cc_{\rm max}=0.275$.

We note that an empirical formula
\begin{equation}\label{empirical}
    a=2(1+0.677\cc)/5
\end{equation} has been proposed previously by fitting the
numerical data of the moment of inertia of QSs constructed with
several different quark matter EOS, which are basically the MIT
bag model (\ref{QMEOS}) plus some minor corrections
\cite{SS,MomentofInertia}. Comparing the
 analytic result in (\ref{Amoi7}), namely $a=2(
1+0.686\cc+0.625\cc^{2}+\cdots)/5$, with the empirical one, we see
that  the empirical formula (shown by the dashed straight line in
Fig.~\ref{ac-curve}) is a good approximation at low compactness
because to order $\cc$ it is almost identical to our
 analytical result. However, as shown in Fig.~\ref{ac-curve}, it
fails to follow the nonlinear behavior of the normalized moment of
inertia for QSs with large compactness, say, $\cc >0.2$,
reflecting the importance of the higher order terms in
(\ref{Amoi7}) in such situation.

\begin{figure}
\centering \caption{The values of the Love number $k_2$ obtained
from the sixth-order post-Minkowsian expansion shown in
\eqref{QS-k2-app} (the continuous solid line) and numerical
integration (the dots), respectively, are plotted against
compactness $\cc$ and compared with each other. }\label{k2c-curve}
\includegraphics[scale=0.5]{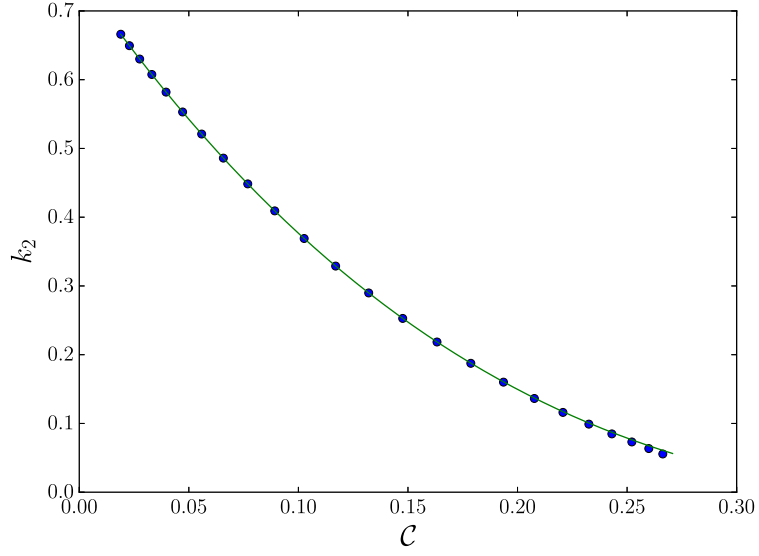}
\end{figure}

Lastly, we express the tidal Love number $k_2$ in terms of the
sixth-order post-Minkowsian expansion
\begin{equation}
k_2=(1-2\cc)^2\left\{\frac{3}{4}-\frac{45 {\cal C}}{28}+\frac{187
   {\cal C}^2}{2940}-\frac{1141589 {\cal C}^3}{1131900}-\frac{273911383
   {\cal C}^4}{103002900}-\frac{374294273707
   {\cal C}^5}{54076522500}
   -\frac{6469553810716163
   {\cal C}^6}{353930839762500}+\cdots\right\},
   \label{QS-k2-app}
\end{equation}
and accordingly a plot of $k_2$ versus $\cc$ is given in
Fig.~\ref{k2c-curve}. Again we see that the agreement between the
numerical and perturbative results is nice.

\newcommand{\noopsort}[1]{} \newcommand{\printfirst}[2]{#1}
  \newcommand{\singleletter}[1]{#1} \newcommand{\switchargs}[2]{#2#1}

\clearpage

\newpage

\end{document}